\documentclass[preprint]{JHEP3} 

\JHEPspecialurl{http://jhep.sissa.it/JOURNAL/JHEP3.tar.gz}

\usepackage{epsfig,multicol}

\def\beq{\begin{equation}}

\def\eeq{\end{equation}}

\newcommand{\bea}{\begin{eqnarray}}

\newcommand{\eea}{\end{eqnarray}}

\parskip=\itemsep               

\newcommand{\beqar}[1]{\begin{eqnarray}\label{#1}}

\newcommand{\eeqar}{\end{eqnarray}}

\def\thefootnote{\fnsymbol{footnote}} 

\title{
{\Large  \bf QCD Reggeon Field Theory for every day: Pomeron loops included}}
\author{Tolga Altinoluk${}^{\,1}$, Alex Kovner${}^{\,1}$, Michael Lublinsky${}^{\,2}$ 
 and Javier Peressutti${}^{\,1}$\\
 ${}^{\,1}$Physics Department, University of Connecticut, 2152 Hillside road, Storrs, CT 06269, USA\\
${}^{\,2}$Physics Department, State University of New York, Stony Brook, NY 11794, USA}

\abstract{We derive the evolution equation for hadronic scattering amplitude at high energy. Our derivation 
includes the nonlinear effects of finite partonic density in the hadronic wave function as well as the effect of multiple scatterings for scattering on dense hadronic target. It thus includes Pomeron loops. It
is based on the evolution of the hadronic wave function derived in \cite{foam}. The kernel of the evolution equation defines the second quantized Hamiltonian of the QCD Reggeon Field Theory, $H_{RFT}$ beyond the limits considered so far. The two previously known limits of the evolution: dilute target (JIMWLK limit) and dilute projectile (KLWMIJ limit) are recovered directly from our final result. The Hamiltonian $H_{RFT}$ is applicable for the evolution of scattering amplitude for arbitrarily dense hadronic projectiles/targets - from "dipole-dipole" to "nucleus-nucleus" scattering processes.
\\

}

\begin{document}



\def\thefootnote{\arabic{footnote}} 


\section{Introduction}
This paper continues our investigation of high energy evolution of hadronic cross sections and other observables. The subject has a long history, starting with Gribov's ideas of reggeon field theory \cite{gribov} and early work exploring Pomeron interactions \cite{Amati}, \cite{Baker}. In the framework of QCD the perturbative evolution equation - the BFKL equation was derived in \cite{BFKL}. It describes the evolution of forward and non-forward scattering amplitudes as the energy of the collision becomes large. The BFKL equation was clearly a crucial milestone in the study of high energy scattering. It has given an impetus to a lot of theoretical and experimental work. However it has been realized very quickly that at very high energies the BFKL equation violates unitarity and leads to cross sections which rise as a power of energy. This violation is very severe, since it does not only violate the Froissart bound for the total cross section, but also violates unitarity of the 
scattering amplitude at fixed impact parameter. In order to restore unitarity one has to include 
$t$-channel exchanges with more than just two reggeized gluons. This idea has been developed  by Bartels \cite{ELLA}. Within this framework the generalization of the BFKL equation is the BKP
equation, which governs (perturbative) high energy evolution of an amplitude due to an exchange of an arbitrary but fixed number of reggeized gluons
\cite{BKP}.

It was further realized that to achieve the $s$-channel unitarity one has to introduce 
transitions between states with different number of reggeized gluons \cite{3P}.
The ideas put forward in \cite{3P} have been under intense investigation during the last decade or so \cite{BLW,Lotter,BV,BBV,Braun1,Braun2}.  At present,
this approach provides elements of an effective theory in terms of  $t$-channel gluon states and transition vertices.
These elements have been put together in \cite{Braun2} into an
 effective theory of BFKL Pomerons interacting via triple pomeron vertex of \cite{3P}.
This model is meant to describe nucleus-nucleus collisions at high energies at LLA and large $N_c$. 
We also note that Lipatov
and collaborators \cite{LipatovFT} have derived an effective action with both real and reggeized gluons as effective degrees of freedom. This action  respects 
the unitarity of full QCD, but its complexity has so far precluded any progress in understanding its physical consequences.

 It is hoped that the end result of this direction of research will be a quantum theory of interacting Reggeons derived entirely from QCD.

A parallel line of research originated by  Gribov, Levin and Ryskin \cite{GLR} has been vigorously pursued in the last 15 years or so. It is based on the idea of gluon saturation. At high energy the evolution of physical observables should slow down because of nonlinear effects due to large density of gluons in the hadronic wave function.
The GLR equation - the nonlinear evolution equation for gluon density in the double logarithmic approximation was derived
in \cite{GLR} and put on firmer theoretical grounds in \cite{MUQI}. 
The gluon saturation ideas have been further developed in a series of papers by Mueller\cite{Mueller}, who also introduced the notion of QCD dipoles as a convenient basis for the discussion of high energy processes, and has related it to BFKL Pomeron and the triple Pomeron vertex \cite{reggeon}.

It was noted in \cite{mv} that the problem of saturation can be also related to nonlinearities in the classical Yang-Mills equations. Following this observation, the formal path integral approach to the problem of evolution has been developed in  \cite{JIMWLK}. This together with 
independent approaches of \cite{balitsky} and \cite{Kovchegov}, and a later work \cite{cgc} resulted in the derivation of a functional evolution equation for the correlators of the color charge density in the hadronic wave function, the so called JIMWLK equation (also called sometimes the Balitsky hierarchy). The equations are more conveniently written as evolution of the correlators of Wilson lines, introduced in  \cite{Verlinde} as effective degrees of freedom at high energy. The JIMWLK equation takes into account the effects of nonlinearities in the wave function of the projectile, but when applied to the scattering amplitude, it does not account for the bulk of multiple scattering corrections. It is applicable in a situation when a dense object ("nucleus") scatters off a dilute perturbative object ("dipole"). The equation therefore does not include the Pomeron loop effects
\cite{MS,IM,KL,shoshi1,IT}. The reverse situation - that of a perturbative "dipole" scattering off a dense "nucleus" was considered in \cite{klwmij} and the evolution equation for this limit has been derived. We will refer to this as the KLWMIJ equation \cite{klwmij}. Although it describes the same physical process as JIMWLK, the KLWMIJ evolution acts on the wave function of a dilute object, and thus does not include the effects of nonlinearities. It does, on the other hand, take into account all important multiple scattering effects. 
The efforts to consistently include Pomeron loops into the evolution have continued since, yielding many interesting developments \cite{MSW,LL3,kl1,something,kl4,kl5,SMITH,MMSW,HIMS,genya,Balitsky05,nonlin,zakopane}.
Also a statistical analogy of the Pomeron loop effects has been suggested in \cite{stat,stat2}, although its validity has not been convincingly demonstrated.

Other recent interesting developments in this area  include the calculation of the next to leading order corrections to the JIMWLK kernel \cite{nexttoleading, KWnlo} as well as the generalization of the approach to calculate less inclusive observables than the scattering amplitude \cite{kovtuch, multigJK,multig, diff}.

The last years also have seen progress in relating the JIMWLK approach to the reggeon field theory.
This relation has been discussed in some detail in \cite{seven}. The functional JIMWLK equation has direct interpretation as the Schroedinger equation of the Hamiltonian reggeon field theory. 
 The JIMWLK kernel in this view is simply the Hamiltonian of the RFT (Reggeon field theory).
 To be precise, the theory which we dub RFT is the effective theory of QCD at high energies  which governs the high energy behavior of scattering amplitudes  of hadrons.
The degrees of freedom of this effective theory are the Wilson lines (single gluon scattering amplitudes). These are not quite the reggeized gluons of \cite{ELLA},\cite{BKP} but are closely related to them.  The mapping between the Wilson lines  and reggeized gluons of
the standard formulation has been discussed in detail in \cite{seven}. Since JIMWLK equation does not contain the Pomeron loop effects, the RFT Hamiltonian obtained in this way is not complete.  To obtain the complete RFT Hamiltonian one has to take into account both the effects of the nonlinear corrections in the projectile wave function as well as contributions of multiple scattering corrections to the scattering amplitude, in the same framework.  Although  the complete RFT Hamiltonian has not been derived yet, some of its general properties have been discussed in \cite{yinyang}. In particular, since  $H_{RFT}$ generates the energy evolution of hadronic scattering amplitudes, unitarity requires the spectrum of its eigenvalues to be positive.

An important step towards the derivation of the complete $H_{RFT}$ was made in \cite{foam}. This paper derived the evolution of the hadronic wave function under boost. The one step that is still missing in \cite{foam} is the derivation of $H_{RFT}$ itself. To derive $H_{RFT}$ one has to consider calculation of physical observables, such as the scattering amplitude in the evolved wave function, rather than simply the evolution of the wave function itself. 

The main purpose of the present paper is precisely to make this additional step and complete the derivation of $H_{RFT}$. Our final result for the RFT Hamiltonian is given in eq.(\ref{hg}). This result is valid to leading order in the coupling constant $\alpha_s$
at any parametric value of the color charge density of the colliding hadrons. It also accounts for all multiple scattering effects in the eikonal approximation. The term eikonal is used here only in the sense that individual partons in the hadronic wave function scatter eikonally. The hadron as a whole in this approximation does undergo both elastic and inelastic scattering \cite{eikonalnot}. This approximation is the same as used in all the recent papers on the subject quoted above.

The Hamiltonian $H_{RFT}$ effectively resums all perturbative 
diagrams in which every factor $\alpha_s$ is enhanced by a single logarithm of energy 
(leading log approximation). In all regimes explored previously, in addition to the leading log 
approximation some assumption about partonic densities was made. The density is either considered to be parametrically
small (dilute regime - KLWMIJ evolution) or parametrically large (dense regime - JIMWLK evolution). Although we will be frequently
referring to the dense regime in this paper, the Hamiltonian we derive is valid for arbitrary density and thus
has only one resummation parameter $\alpha_s\,Y\,\simeq\,1$. Corrections to our result
are suppressed by a power of $\alpha_s$ at any partonic density. Although we have not studied these corrections in full generality, one can easily see that in both the JIMWLK and the KLWMIJ limits the expansion is in integer powers of $\alpha_s$ and 
any correction to our result contributes at most
at next-to-leading order in this expansion.

The result presented in eq.(\ref{hg}) generalizes all previously available limiting expressions for $H_{RFT}$ \cite{JIMWLK, cgc, Kovchegov, balitsky, klwmij, kl1, SMITH, Balitsky05} to the most general situation. It reproduces the JIMWLK and the KLWMIJ Hamiltonians in the appropriate limits, and is applicable also when both colliding objects are either small ("dipole-dipole scattering") or large ("nucleus-nucleus scattering").

Similarly to previously known limits, 
the complete $H_{RFT}$ eq.(\ref{hg}) can be expressed in terms of  Wilson lines. Those are $SU(N)$ unitary matrices  which depend
on the two dimensional transverse coordinates. These are the basics quantum fields - the degrees of freedom of the effective Reggeon Field Theory. 
We note that one can introduce
two sets of Wilson lines (associated to the propagation of projectile or target partons), which play the role of (almost) canonically conjugate variables. Either set of the Wilson lines can be chosen as a complete set of variables on the Hilbert space of RFT.

The scattering amplitude is not the only physical observable whose evolution is governed by $H_{RFT}$. In the companion paper \cite{Q} we continue this line of investigation by deriving expressions for more exclusive observables, which also resum effects of Pomeron loops. In particular we discuss in \cite{Q} the single gluon inclusive spectrum and also the double and multi gluon exclusive spectra when none of the observed gluons
are separated by large rapidity, so that the effects of energy evolution between the final state gluon rapidities are unimportant.

This paper is structured as follows. In Section 2 we recap the main results of \cite{foam} and outline the steps for calculation of $H_{RFT}$. Section 3 contains the calculations of main ingredients necessary for the final push. In Section 4 we complete the calculation  of the $H_{RFT}$ and derive our main result. We discuss our results in Section 5. Appendices provide some details of calculation as well as discuss some aspects of the perturbative expansion of our result.

\section{The Summary and the Road Map.}

In \cite{foam} we have analyzed the evolution of the hadronic wave function at high energy. Here we summarize the results of \cite{foam}.

The standard expansion of the field operator in terms of creation and 
annihilation operators of free gluons is
\begin{equation}
A^a_i(x^-,x)\ =\ \int_0^\infty {dk^+\over 2\pi}\,{d^2k\over 4\pi^2}\, {1\over \sqrt {2k^+}}
\left\{a^a_i(k^+,k)\,e^{-ik^+x^-\,-ikx}\ +\ a^{a\dagger}_i(k^+,k)\,e^{ik^+x^-\,+ikx}\right\}
\end{equation}
where
\begin{equation}
[a^a_i(k^+,k),a^{\dagger b}_j(p^+,p)]\,=\,(2\pi)^3\delta(k^+-p^+)\,\delta^2(k-p)\,.
\end{equation}
The operator $A^a_i(x^-,x)$ is the vector potential operator in the light-cone gauge $A^+=0$ with the zero longitudinal momentum mode subtracted.\footnote{In \cite{foam} it was denoted as $\tilde A$.}
Throughout the text we will be using upper Latin indices $a,b\ldots=1,...,N^2-1$ to denote color while
lower indices $i,j,\ldots=1,2$ to denote rotational components.
For further reference we note that the vector potential can also be decomposed in the creation and annihilation operators in the frequency basis:
\begin{equation}
A^a_i(x^-,x)\,=\,\int_0^\infty {dk^-\over 2\pi}\,{d^2k\over 4\pi^2}\, {1\over \sqrt {2k^-}}\,
\left\{a^a_i(k^-,k)\,e^{-i{k^2\over 2k^-}x^-\,-ikx}\,+\,a^{a\dagger}_i(k^-,k)\,e^{i{k^2\over 2k^-}x^-\,+ikx}\right\}
\end{equation}
where
\begin{equation}
a^a_i(k^-,k)\,=\,\sqrt{k^+\over k^-}\,a^a_i(k^+,k)
\end{equation}
so that 
\begin{equation}
[a^a_i(k^-,k),a^{\dagger b}_j(p^-,p)]\,=\,(2\pi)^3\,\delta(k^--p^-)\,\delta^2(k-p)\,.
\end{equation}

We consider a hadronic projectile moving to the right with large energy.
Suppose at initial rapidity we know the hadronic wave function $|\Psi\rangle$. 
The bulk of gluons in this wave function are at rapidities greater than some "cut off" rapidity.
 The properties of this valence component of the wave function are characterized by correlators of the color charge density operator $j^a(x)$.  The gluons with rapidity smaller
than the cutoff  are small in number and therefore do not contribute to observables in the leading order in the coupling constant. 

This "soft" component of the wave function can be calculated perturbatively.
It was shown in \cite{foam} that the  wave function, including its soft component 
has the form
\begin{equation}\label{psi}
|\Psi\rangle\,=\,\Omega[a,a^\dagger,j]\,|v\rangle
\end{equation}
where $a$ and $a^\dagger$ are soft gluon creation and annihilation operators with rapidities below the cutoff.
The valence state $|v\rangle$ has no soft gluons and is therefore annihilated by the soft gluon annihilation operators
$$
a\,|v\rangle\,=\,0\,.
$$
The evolution operator $\Omega$ is a unitary operator of the Bogoliubov type
\begin{equation}\label{prod}
\Omega\,=\,{\cal C}\,{\cal B}\,.
\end{equation}
Here $\cal C$ is a coherent operator that creates the "classical" Weizsaker-Williams field
\begin{equation}\label{coherent}
{\cal C}\,=\,\exp\left\{2i\int\, d^2x\,b^a_i(x)\, A^a_i(x^-=0,x)\right\}\, .
\end{equation}
and $\cal B$ is a Bogoliubov type operator responsible for the leading quantum corrections (see below).

The color charge operators $j(x)$ are operators on the valence Hilbert space and form a local $SU(N)$ algebra:
\beq
[j^a(x),\,j^b(y)]\,=\,i\,g\,f^{abc}\,j^c(x)\,\delta^2(x-y)\,.
\eeq
The Weizsaker-Williams field $b_i^a(x)$ created by the valence modes
depends only on transverse coordinates $x$ and is a two dimensional pure gauge field. It can therefore be written in terms of a  unitary matrix $U$ as
\begin{equation}
b^a_i\,=\,-{1\over g}\,f^{abc}U^{\dagger bd}\,\partial_i \,U^{dc}
\end{equation}
where $f^{abc}$ are the structure constants of the $SU(N)$ and $U^{ab}$ is an $SU(N)$ group element in the adjoint representation. The Weizsaker-Williams field is related to the valence color charge density by
\begin{equation}
\partial_i\,b^a_i(x)\,=\,j^a(x)
\end{equation}
The classical field $b$ has a well known diagrammatic interpretation. It sums the tree level diagrams to the one gluon component of the state created by the coherent operator ${\cal C}$ from the Fock vacuum (see Fig. 1).
\FIGURE{\epsfig{file=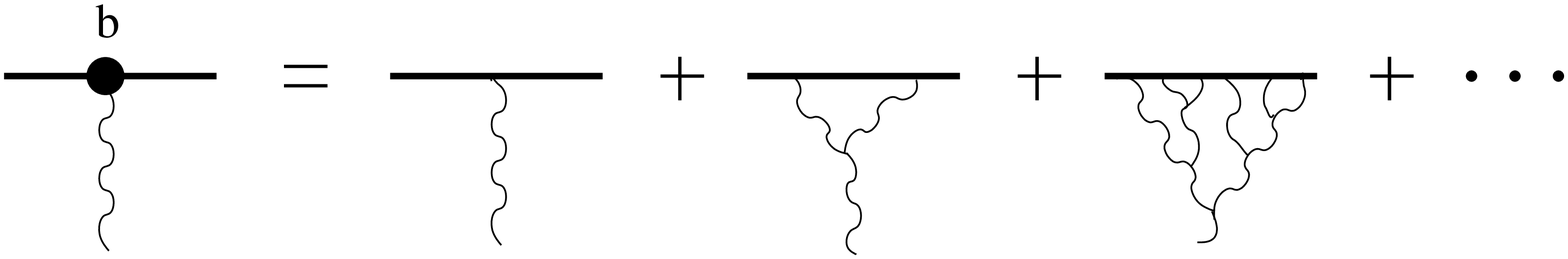,width=120mm}
\caption{\it Diagrammatics of the classical field $b^a_i$. The straight solid line represents the valence charges which serve as sources of the classical field.}
\label{fig1}
}

The explicit form of $\cal B$ has not been given in \cite{foam}, however the action of $\cal B$ on the gluon fields was calculated. Since we will use heavily this in the following, we give here the relevant expressions, even though they are quite lengthy.
\begin{equation}\label{transformed}
{\cal B}^\dagger\,A^a_i(x)\,{\cal B}\,=\,c^a_i(x)+\epsilon(x^-)\Delta^{ab}_{ij}(x,y)c^b_j(y^-=0,y)\, ,
\end{equation}
where
\begin{equation}
\epsilon(x)\,=\,\frac{1}{2}\left[\theta(x)- \theta(-x)\right]\nonumber
\end{equation}
and
\begin{eqnarray}
c_{i}^a(x)&=&\int_{0}^\infty \frac{dp^-}{ 2\pi}\int d^2q \left[\theta(-x^-)e^{i\frac{\partial^2}{2p^-}x^-}[t-l]_{ij}(x,y)v^{aj}_{p^-,q}(y)\right.
\nonumber \\
&& \left. \qquad \qquad \qquad \qquad +\theta(x^-)e^{i\frac{D^2}{2p^-}x^-}[T-L]^{ab}_{ij}(x,y)v^{bj}_{p^-,q}(y)\right]a_{p^-,q}+h.c. \, ,
\label{canonical}
\end{eqnarray}
with
\begin{equation}
\Delta^{ab}_{ij}(x,y)=\left\{D_i\frac{1}{ \partial D}D_j+D_i\frac{1}{D\partial }D_j-2D_i\frac{1}{\partial D}\partial_j\right\}^{ab}(x,y)\, .
\end{equation}
The covariant derivative is $D^{bc}_i=\partial_i\delta^{bc}-f^a_{bc}b^a_i$
and the projectors $T,\ L, t,\ l$ are defined as
\begin{equation}
L^{ab}_{ij}=\left[D_i\frac{1}{ D^2}D_j\right]^{ab},\ \ \ \ \ \ 
\ T^{ab}_{ij}=\delta^{ab}_{ij}-L^{ab}_{ij}; \ \ \ \ \ \  l_{ij}=\partial_i\frac{1}{\partial^2}\partial_j; \ \ \
\ \ \ \ t_{ij}=\delta_{ij}-l_{ij}\, .
\end{equation}
The integral over $p^-$ in eq.(\ref{canonical}) strictly speaking excludes the point $p^-=0$ since  soft gluon modes all have non-vanishing light cone frequency. This subtlety however does
not affect any of our calculations and we will therefore not indicate this explicitly.
 
The transverse basis functions $v_{p^-q}$ are the analogs of plane waves and are normalized according to
\begin{equation}
\int d^2q\,v_{p^-q}^{-i}(x)\,v_{p^-q}^{*-j}(y)\,=\,{\cal W}^{ij}_{p^-}(x,y)\, .
\end{equation}
where
\begin{equation}\label{M}
{\cal W}^{ij}_{p^-}(x,y)=\left(\frac{1}{p^-}\right)^2\left\{\frac{1}{\frac{1}{p^-}+i\epsilon}[\delta^{ij}\delta^2(x-y)+\frac{1}{2}C^{ij}(x,y)]+\frac{1}{\frac{1}{p^-}-i\epsilon}[\delta^{ij}\delta^2(x-y)-\frac{1}{2}C^{ij}(x,y)]\right\}\, .
\end{equation}
with
\begin{equation}
C^{ab}_{ij}(x,y)=\left\{2\partial_i\frac{1}{D\partial }D_j-2D_i\frac{1}{\partial D}\partial_j\right\}^{ab}(x,y)\, .
\label{2.29}
\end{equation}
For all practical purposes this means that for any finite frequency $p^-$ the functions $v$ can be taken as plane waves (we suppress rotational and color indices in which $v$ can be taken as a unit matrix)
\begin{equation}
v_{p^-,q}(x)\,=\,{1\over \sqrt {p^-}}\,e^{iqx}\,.
\end{equation}
For the mode at infinite frequency the normalization is different as is given by the operator 
$C_{ij}^{ab}(x)$. Although the infinite frequency mode does not contribute directly to any quantity which is proportional to rapidity, it does play an important role in insuring the completeness of the expansion basis in eq.(\ref{transformed}). We will not be working with this mode directly, but will simply use the fact that including this mode, the operator $\cal B$ is indeed unitary, with all the ensuing consequences.

Since the color charge density is a quantum operator on the valence Hilbert space, it is also transformed by the action of $\cal B$. This transformation has been found explicitly in \cite{foam}. It turns out that for our present purposes it is not necessary to know the exact form of this transformation apart from the fact that it adds to $j$ terms of relative order $O(\alpha_s)$.  

We now consider the energy evolution of the hadronic wave function.
Increasing energy of a hadron is equivalent to boosting it. Under the boost transformation, the longitudinal momenta of the soft gluons in the wave function eq.(\ref{psi}) increase. As a result, soft gluons emerge from below the "cutoff"  and
the number of gluons in the wave function which contribute to any physical observable, such as a scattering amplitude increases  (Fig. 2). 
\FIGURE{\epsfig{file=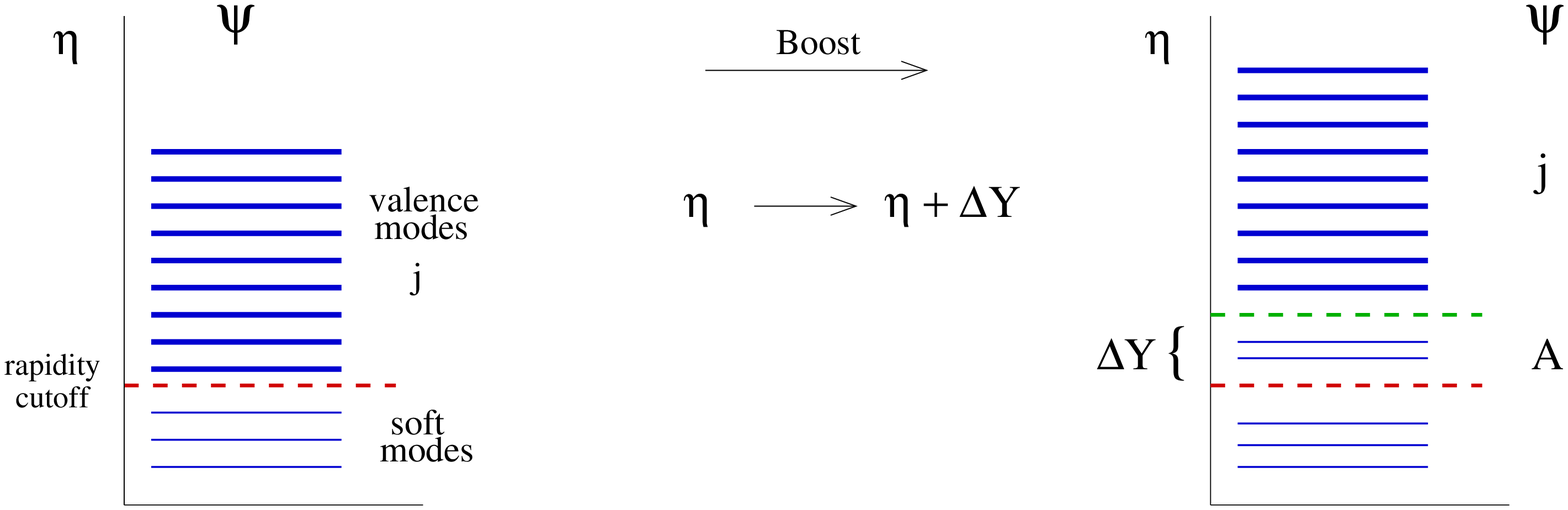,width=140mm}
\caption{\it The light cone wave function before and after boost.
 Here $\eta$ stands for gluon`s rapidity, $j$ for the color charge density of the valence gluons, and $A$ for the soft gluon field.}
\label{fig2}}

Thus in the leading logarithmic approximation the soft part of the wave function determines completely the evolution of physical observables with rapidity. 
However to find this evolution one has to make an additional step, and consider directly some physical observable. The purpose of this paper is precisely to make this additional step from the soft gluon wave function to physical observables.

The most basic and most inclusive observable is the forward scattering amplitude of our projectile hadron on some hadronic target. This is the observable we are going to discuss in this paper.
It belongs to a broader class of observables which depend only on the correlators of the color charge density operator $j^a(x)$ and not on other characteristics of the projectile wave function. Evolution of observables of this type is governed by the functional evolution equation which generalizes the JIMWLK \cite{JIMWLK,cgc} and KLWMIJ \cite{klwmij} - the Schroedinger equation of RFT.  The kernel of the evolution equation can be viewed as the second quantized
Hamiltonian of the QCD Reggeon Field Theory (RFT) as discussed at length in \cite{seven}. By considering an arbitrarily dense hadron as a possible target in the scattering process, we will derive the functional evolution equation which is applicable in the full parameter range between the BFKL scattering of two small perturbative objects through to nucleus-nucleus scattering. The result of the present paper is therefore the RFT Hamiltonian which contains the effects of Pomeron loops.

Another example of observables of this type are 
 the single inclusive, double inclusive and in general multiple gluon production amplitudes, where all gluons are close to each other in rapidity.  In the companion paper \cite{Q} we derive the explicit expressions for these observables in terms of the target and projectile fields, to be averaged over the corresponding wave functions, and discuss their evolution. 

We now briefly sketch what we have to do to derive $H_{RFT}$.
Consider the calculation of some hadronic observable $O$ which depends only on the color charge density operator $j^a(x)$ and not on any other property of the wave function
\begin{equation}\label{O}
\bar O\,=\,\langle v\vert\, \hat O[j]\,\vert v\rangle\,=\,\int Dj\,W^P[j]\,O[j]\,.
\end{equation}
The last expression is by now the standard way of representing the wavefunction average using functional integral. The weight functional
$W^P$ provides a probability distribution for the valence charges $j$ in the wavefunction. 

An important (but not the only) example of an observable discussed above is the eikonal $S$ - matrix in the external field $\alpha$
\begin{equation}\label{eikonal}
\hat S\,=\,\exp\left\{i\int_x j^a(x)\,\alpha^a(x)\right\}\,.
\end{equation}
In the approach of \cite{JIMWLK} the expression of the scattering amplitude for a hadron-hadron scattering is given by
\begin{equation}
{\cal S}\,=\,\langle\langle v|\,\hat S\,|v\rangle\rangle_{\alpha}
\end{equation}
where the weight for the averaging over the fields $\alpha$ is determined by the  wavefunction of the target.

The same procedure applies to any observable.
The matrix element $\bar O$ is  the
expectation value of the operator $O$ in the projectile wavefunction. In order to
obtain the final expression for the observable measured in the collision, we also have to average $\bar O$ over the target wavefunction.
\begin{equation}\label{Ot}
\langle\,\bar O\,\rangle\,_T\,=\,\int D\alpha\,W^T[\alpha]\,\,\bar O
\end{equation}
where $\alpha$ represents all the target fields\footnote{In most of our discussion
below we will be focused on the evolution of the projectile and will not indicate target averages explicitly. One should remember, however, that 
the averaging over the target (\ref{Ot}) is always assumed in final results.}.

We assume, as before that only valence degrees of freedom contribute to the average eq.(\ref{O}) at initial rapidity. When the system is boosted to rapidity $Y$, two important changes occur. First, the relevant wave function changes from $|v\rangle$ to $|\Psi\rangle$, and second the color charge density of the soft gluons has to be added to $j^a$ in the observable $O[j]$. Both changes are due to the fact that the soft gluons after boost emerge with momenta above the cutoff rapidity, and now contribute to physical observables (see Fig. 2).
Thus eq.(\ref{O}) at higher rapidity $Y$ becomes 
\begin{equation}\label{oy}
\bar O_Y\,=\,
\langle \Psi\vert \  O[j^a(x)+j^a_{soft}(x)]\ \vert \Psi\rangle\ =\ \langle v|\ \Omega^\dagger \,\hat R_a \, O[j]\ \Omega\ |v\rangle\,.
\end{equation}
Here the charge density shift operator 
\begin{equation}\label{R}
\hat R_a\,\equiv\, e^{\ \int_x j^a_{soft}(x)\ {\delta\over \delta j^a(x)}}
\end{equation}
and
\begin{equation}
j^a_{soft}(x)\,=\,g\,\int {dk^+\over 2\pi}\, a^{\dagger\,b}(k^+,x) \ T^a_{bc}\ a^c(k^+,x)
\end{equation}
is the color charge density of the soft gluons with the integral over $k^+$ defined over the range of momenta corresponding to the boost parameter $\Delta Y$ (Fig. 2).

Eq.(\ref{oy}) has a universal form, in the sense that it does not depend on the particular operator $O$.
Since the state $|v\rangle$ is the vacuum of the soft gluon degrees of freedom, it is possible in principle to calculate the expectation value in eq.(\ref{oy}) over the soft gluon part of the Hilbert space without the explicit knowledge of the operator $ O$. This involves calculating the following matrix element
\begin{equation}\label{exp}
\langle 0|\,\Omega^\dagger[j,a,a^\dagger] \ e^{\int_x \,j^a_{soft}(x)\,{\delta\over \delta j^a(x)}}\ \Omega[j,a,a^\dagger]\,|0\rangle\,.
\end{equation}
The quantum averages in eq.(\ref{exp}) are performed only over the soft gluon Hilbert space.
Here the functional derivatives $\delta/\delta j$ are treated as $c$-numbers since according to eq.(\ref{oy}) they should only act on 
 $j$'s in the operator $O$. 
The valence color charges $j^a$ in this expression are "almost" c-numbers, in the sense that they are not quantum operators on the Hilbert space of the soft gluons. One should however keep track of their ordering, since they do not commute with each other.

The expression in eq.(\ref{exp}) of course depends on the total rapidity $\Delta Y$
 available to gluons in the soft Hilbert space. Since we are interested in deriving a differential equation in this rapidity, we have to treat this rapidity as small. To obtain the kernel of the high energy evolution equation we simply have to calculate eq.(\ref{exp}) to first order in this total rapidity. 
\beq
{d\over dY}\bar O\,=\,\lim_{\Delta Y\rightarrow 0}\frac{\langle v|\ \Omega^\dagger \,\hat R_a \,\hat O[j]\ \Omega\ |v\rangle\ -\ 
\langle v|\, O[j]\, |v\rangle}{\Delta Y}\ \equiv\ -\,H_{RFT}\left[j,{\delta\over \delta j}\right]\ \langle v|\, O[j]\, |v\rangle
\eeq
where $H_{RFT}$ is defined as
\begin{equation}\label{defin}
H_{RFT}\left[j,{\delta\over \delta j}\right]\,=\,-{d\over dY}\langle 0|\Omega^\dagger[j,a,a^\dagger] \,
\hat R_a\, \Omega[j,a,a^\dagger]|0\rangle|_{Y=0}\,.
\end{equation}
We see that the evolution of any observable $\bar O$ is given by
\begin{equation}
{d\over dY}\bar O\ =\ -\,\langle v|\ H_{RFT}\left[j,{\delta\over \delta j}\right]\ O[j]\ |v\rangle\ =\ -
\int Dj\  W^P[j]\ H_{RFT}\left[j,{\delta\over \delta j}\right]\ O[j]\,.
\end{equation}
Conforming with literature, in the last line we have represented averaging over the valence state as a functional integral over the color charge density with the weight functional $W^P[j]$.
The functional derivatives can be integrated by parts to make them act on $W^P$.
Assuming that the Hamiltonian $H_{RFT}$ is Hermitian (which we will find to be the case), 
we can rewrite this equation as an evolution equation for the weight functional $W^P[j]$ as
\begin{equation}
{d\over dY}\,W_Y^P[j]\ =\ -\,H_{RFT}\left[j,{\delta\over \delta j}\right]\ W_Y^P[j]\,.
\end{equation}

Thus we conclude that in order to find the evolution of the scattering amplitude and other observables which depend only on the color charge density,  we need to calculate the matrix element in eq.(\ref{exp}) and expand it to first order in $Y$.
This is what we will do in the next section.

\section{The Derivation}

We start by deriving a more explicit expression for the wave function
\begin{equation}
|\Psi\rangle\ =\ \Omega[j,a,a^\dagger]\ |0\rangle\,.
\end{equation}
We find it convenient to define the creation and annihilation operators in rapidity basis rather than in the basis of longitudinal momentum 
$k^+$ or frequency $k^-$ as
described in the introduction. 
 Defining the rapidity variable as $\eta=\ln{p^-_0\over p^-}$ with an arbitrary constant $p^-_0$, we  rescale the creation and annihilation operators as
\begin{eqnarray}
&&a^a_i(k^+,k)\,=\,\sqrt{1\over k^+}\,a^a_i(\eta,k)\\
&&[a^a_i(\eta,k),a^{\dagger b}_j(\xi,p)]\,=\,(2\pi)^3\,\delta(\eta-\xi)\,\delta^2(k-p)\,.\nonumber
\end{eqnarray}
In terms of these operators we have
\begin{equation}\label{arap}
A^a_i(x^-,x)={1\over \sqrt 2}\int_{-\infty}^\infty {d\eta\over 2\pi}{d^2k\over 4\pi^2}
\left\{f^{(b,j,\eta,k)}(a,i,x^-,x)a^b_j(\eta,k)+f^{*(b,j,\eta,k)}(a,i,x^-,x)a^{b\dagger}_j(\eta,k)\right\}
\end{equation}
with
\begin{equation}\label{fs}
f^{(a,i,\eta,k)}(b,j,x^-,x)\,=\,\delta^{ab}\,\delta_{ij}\,e^{-i\,{k^2\over 2p^-_0}\,e^\eta x^-\,-\,ikx}\,.
\end{equation}
Strictly speaking the soft modes $a(\eta)$ live only at rapidities below  
$Y$, which is given by the parameter of the boost from $|v\rangle$ to $|\Psi\rangle$. 
We do not denote this cutoff explicitly in most of our formulae and 
extend the  rapidity integration over $\eta$ to infinity for all quantities for which the integration converges. 
We will indicate the $Y$ dependence for divergent quantities and eventually this very dependence will determine the evolution in rapidity.

The coherent state operator in rapidity basis is rather simple
\begin{equation}\label{cohrap}
{\cal C}\,=\,\exp\left\{\,i\,\sqrt 2\,\int {d^2k\over (2\pi)^2}\ b_i^a(k)\ \int {d\eta\over 2\pi}\ 
[a^a_i(\eta,k)\,+\,a^{\dagger a}_i(\eta,-k)]\right\}\,.
\end{equation}
\subsection{The vacuum of $\beta$}

Given that $\Omega$ is a product as in eq.(\ref{prod}), we first analyze the state
\begin{equation}
|0\rangle_\beta\,\equiv\,{\cal B}\,|0\rangle\,.
\end{equation}
Clearly, the state $|0\rangle_\beta$ is annihilated by the operator 
$\beta_\alpha\ =\ {\cal B}\, a_\alpha \, {\cal B}^\dagger\,.$
Here and in the following we denote all the indices of operators $a$, $a^\dagger$ etc by a single Greek  letter $\alpha$. This includes rotational and color indices, as well as the transverse momentum and rapidity\footnote{Below we will be frequently suppressing this index and using the more compact matrix notations instead.}.
Since the operator $\cal B$ is of the Bogoliubov type, the transformation between the operators $a,\ \ a^\dagger$ and $\beta,\ \ \beta^\dagger$ is
 linear homogeneous and can be written quite generally as
\begin{equation}
\beta_\alpha\ =\ {\cal B}\, a_\alpha \, {\cal B}^\dagger\,=\,\Theta_{\alpha\beta}\ a_\beta\ +\ \Phi_{\alpha\beta}\ a^\dagger_\beta\,,\ \ 
\ \ \ \ \ \ \ \ \ \ \ \beta^\dagger_\alpha\ =\ \Theta^*_{\alpha\beta}\ a^\dagger_\beta\ +\ \Phi^*_{\alpha\beta}a_\beta\,.
\label{ML}
\end{equation}
Since the transformation eq.(\ref{ML}) is canonical, the transformation matrices $\Theta$ and $\Phi$ satisfy
\begin{equation}\label{ident}
\Theta\,\Phi^T\,-\,\Phi\,\Theta^T\,=\,0,\ \ \ \ \ \ \ \ \ \ \ \ \ \ \ \ \ \Theta\,\Theta^\dagger\,-\,\Phi\,\Phi^\dagger\,=\,1\,.
\end{equation}
The inverse transformation is given by
\begin{equation}
a_\alpha\ =\ \Theta^\dagger_{\alpha\beta}\ \beta_\beta\ -\ \Phi^T_{\alpha\beta}\ \beta^\dagger_\beta,\ \ \ \ \ \ \ \ \ \ \ \ \ 
\ a^\dagger_\alpha\ =\ \Theta^T_{\alpha\beta}\ \beta^\dagger_\beta\ -\ Q^\dagger_{\alpha\beta}\ \beta_\beta\,.
\label{PQ}
\end{equation}
This leads to another set of relations
\begin{equation}\label{ident1}
\Theta^\dagger \,\Theta\,-\,\Phi^T\,\Phi^*\,=\,1\,; \ \ \ \ \ \ \ \ \ \ \ \ \ \Theta^\dagger \,\Phi\,-\,\Phi^T\,\Theta^*\,=\,0\,.
\end{equation}
The explicit form of the transformation matrices $\Theta$ and $\Phi$ can be found from the results of \cite{foam} 
and we provide it later in eq.(\ref{MLN}). For now however we keep the discussion general as the explicit form of $\Theta$ and $\Phi$ is not important until late in the game.

To find the state $|0\rangle_\beta$ we write it in the form
\begin{equation}
|0\rangle_\beta\,=\,F^{-1/2}[\Lambda]\ e^{\,-{1\over 2}\,a^\dagger_\alpha\,\Lambda_{\alpha\beta}\, a^\dagger_\beta}\ |0\rangle\,.
\end{equation}
Imposing the condition
\begin{equation}
\beta\,|0\rangle_\beta\,=\,0
\end{equation}
gives
\begin{equation}
(\Theta\, a+\Phi\, a^\dagger)\ e^{\,-{1\over 2}\,a^\dagger \,\Lambda\, a^\dagger}\ |0\rangle\ =\ (-\Theta\,\Lambda\,+\,\Phi)\,
a^\dagger\,
e^{\,-{1\over 2}\,a^\dagger\,\Lambda\, a^\dagger}\ |0\rangle\ =\ 0\,.
\end{equation}
Thus we find
\begin{equation}\label{Lambda}
\Lambda\ =\ \Theta^{-1}\,\Phi\,.\
\end{equation}
To find the normalization of the state we have to calculate
\begin{equation}
F[\Lambda]\, =\,\langle 0|\ e^{\,-{1\over 2}\,a\,\Lambda^\dagger \, a}\ e^{\,-{1\over 2}\,a^\dagger\,\Lambda\, a^\dagger}\ |0\rangle\,.
\end{equation}
For future use we  calculate in Appendix A a slightly more general expression $F[\Lambda,\bar\Lambda]$ 
\begin{equation}\label{F}
F[\Lambda,\bar\Lambda]\,\equiv\,
\langle 0|\ e^{\,-{1\over 2}\,a\,\bar\Lambda^\dagger \, a}\ e^{\,-{1\over 2}\,a^\dagger\,\Lambda a^\dagger}\ |0\rangle
\ =\ \exp\left\{\,-{1\over 2}\,{\rm Tr}\, \ln(1\,-\,\bar\Lambda^\dagger\,\Lambda)\ \right\}
\end{equation}
with an arbitrary matrix $\bar \Lambda$. In the last expression  the trace is over all the indices. 
Noting that  $F[\Lambda]\,=\,F[\Lambda,\Lambda]$ we finally find
\begin{equation}\label{betavac}
|0\rangle_\beta\ =\ e^{\,{1\over 4}\, {\rm Tr}\, \ln(1\,-\,\Lambda^\dagger\,\Lambda)}\ e^{\,-{1\over 2}\,a^\dagger\,
\Lambda \,a^\dagger}\ |0\rangle
\end{equation}
with $\Lambda$ given  by eq. (\ref{Lambda}).
Just like the classical field $b$, the matrix $\Lambda$ has a diagrammatic representation. It represents the two gluon component of the wave function obtained by action of ${\cal B}$ on the Fock vacuum, and can be represented by a two gluon vertex - Fig. 3.
\FIGURE{\epsfig{file=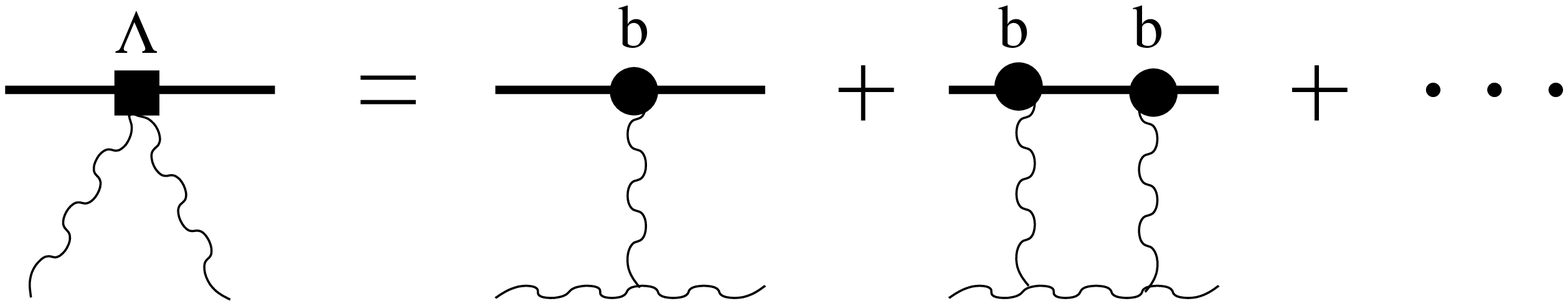,width=100mm}
\caption{\it Diagrammatic representation of the two gluon vertex $\Lambda$. }
\label{fig3}
}
\subsection{The matrix element.}
According to eq.(\ref{defin}) we need to calculate the matrix element of the operator $\hat R_a$. Thus we need to multiply the state $|0\rangle_\beta$ by the coherent operator $\cal C$,  eq.(\ref{coherent}) and then act on it with $\hat R_a$. The operator $\hat R_a$ was introduced as the operator that shifts the valence color charge density by the color charge of the soft gluons. On the other hand when acting on the soft gluon operators it acts as the $SU(N)$ rotation, rotating the soft gluon creation and annihilation operators by the unitary phase
\begin{equation}
\hat R^\dagger_a\, a^b(x)\,\hat R_a\ =\ R^{bc}(x)\,a^c(x), \ \ \ \ \ \ \ \ \ 
\hat R^\dagger_a\, a^{\dagger \,b}(x)\,\hat R_a\ =\ R^{bc}(x)\,a^{\dagger c}(x)
\end{equation}
with 
\begin{equation}
R^{ab}(x)\ =\ \left[{\cal P}\,\exp\{g\int_0^1 dt\  T^c\ {\delta\over\delta j^c(x,t)}\right]^{ab}\,.
\end{equation}


As explained in \cite{klwmij} to properly keep track of the ordering of the non-commuting operators $j^a(x)$ it is convenient to endow the valence color charge density $j$ with the additional "ordering" coordinate $t$; $j(x)\rightarrow j(x,t)$. For the purpose of our present derivation however, the only thing that matters is that $R^{ab}(x)$ is a unitary $c$ - number matrix.

Rotation of $a$ and $a^\dagger$ is equivalent to rotation of both indices of the matrix $\Lambda$ as well as rotation of the classical field $b_i^a(x)$ by the same matrix $R(x)$. Thus we will consider the following general matrix element, which covers the particular case that interests us here
\begin{eqnarray}\label{matel}
G[b\,-\,\bar b]&\equiv& 
\langle 0|\ e^{\,-{1\over 2}\,a_\alpha\,\bar\Lambda^\dagger_{\alpha\beta} \,a_\beta}\ 
e^{\,-i\,\bar b_\alpha\,(a_\alpha\,+\,P_{\alpha\beta}\,a^\dagger_\beta)}\
e^{\,-i\, b_\alpha\,(a_\alpha\,+\,P_{\alpha\beta}\,a^\dagger_\beta)}\ 
e^{\,-{1\over 2}\,a^\dagger_\alpha\,\Lambda_{\alpha\beta}\, a^\dagger_\beta}
\ |0\rangle \ = \nonumber \\
&=&\langle 0|\ e^{\,-{1\over 2}\,a\,\bar\Lambda^\dagger \,a}
\ e^{\,-i \,(b\,-\bar b)\,(a\,+\,P\,a^\dagger)}\ e^{\,-{1\over 2}\,a^\dagger\,\Lambda\, a^\dagger}\ |0\rangle\,.
\end{eqnarray}
So far both the matrix $\bar \Lambda$ and  field $\bar b$ are arbitrary. They will be specified shortly.
We have introduced the matrix $P_{\alpha \beta}$ to be able to reproduce eq.(\ref{cohrap}) in the transverse momentum representation. The structure of the form $\int_{k} b(k)[a(k)+a^\dagger(-k)]$ is reproduced in eq.(\ref{matel}) by choosing the operator $P$ to be diagonal in rapidity, color and rotational indices and to satisfy $P(k,p)\,a(p)=a(-k)$. The operator $P$ is clearly unitary and satisfies $P^2=1$. We will not need to know any additional properties of $P$ in the following.

The calculation is performed in the Appendix A with the result
\begin{eqnarray}\label{solG}
G[b\,-\,\bar b]&=&F[\Lambda,\bar\Lambda]\exp\left\{-{1\over 2}(b-\bar b)(1-\Lambda P)[1-P\bar\Lambda^\dagger\Lambda P]^{-1}
(1-P\bar\Lambda^\dagger)P(b-\bar b)\right\}\\
&=&F[\Lambda,\bar\Lambda]\,\exp\left\{-{1\over 2}(b-\bar b) (1-P\bar\Lambda^\dagger )[1-\Lambda\bar\Lambda^\dagger ]^{-1}
(1-\Lambda P)P(b-\bar b)\right\}\nonumber\,.
\end{eqnarray}

After these preliminaries we are almost ready to write down the answer for the matrix element in eq.(\ref{defin}).
The only additional element we have to take into account is the ordering of the color charge density operators in the operators $\Omega^\dagger$ and $\Omega$. When calculating any physical observable, as in eq.(\ref{oy})
all the color charge density operators in the operator $O$ are ordered such that they are to the right of the operators $j$ in  $\Omega^\dagger$ but to the left of $j$'s in $\Omega$. Thus when we introduce the ordering coordinate $t$, all $j$'s in $\Omega^\dagger$ have to be assigned the value $t=0$, while all $j$'s in the operator $\Omega$ must have value $t=1$.
 In our derivation the fields $\bar b$ and $\bar\Lambda$ are associated with the operator $\Omega^\dagger$ and they depend on $j(x,t=0)$, while the fields $b$ and $\Lambda$ depend on $j(x,t=1)$.  Also, as explained in \cite{kl4}, since any physical weight functional 
$W^P$ depends only on powers of the matrix $R(x)$ acting on $\delta[j]$, all the operators $j$ that appear in the RFT Hamiltonian  should be understood as acting on the matrix $R$ as left or right rotations.
\begin{eqnarray}
j^a(x,t=1)&=&gJ_R^a(x)=-g{\rm tr} \left\{R(x)T^{a}{\delta\over \delta R^\dagger(x)}\right\};\\
j^a(x,t=0)&=&gJ_L^a(x)=
-g{\rm tr} \left\{T^{a}R(x){\delta\over \delta R^\dagger(x)}\right\};\nonumber\\
J_L^a(x)\,\,&=&\,\,[R(x)\,J_R(x)]^a\,.\nonumber
\end{eqnarray}
Thus we define two classical fields, $b_R$ and $b_L$:
\begin{eqnarray}
b^a_{Ri}\,&=&\,-{1\over g}\,f^{abc}U^{\dagger bd}[J_R]\,\partial_i \,U^{dc}[J_R]\equiv\,-{1\over g}\,f^{abc}U_R^{\dagger bd}\,\partial_i \,U_R^{dc}; \nonumber\\
b^a_{Li}\,&=&\,-{1\over g}\,f^{abc}U^{\dagger bd}[J_L]\,\partial_i \,U^{dc}[J_L]\equiv\,-{1\over g}\,f^{abc}U_L^{\dagger bd}\,\partial_i \,U_L^{dc}
\end{eqnarray}
and similarly two matrices $\Lambda$
\begin{equation}
\Lambda_R\equiv
\Lambda[J_R]\,=\,\Theta^{-1}[J_R]\,\Phi[J_R]; \ \ \ \  \ \ \ \ \ \ \ \ \ \ \ \ \ \ \ \ \ \ \ \Lambda_L\equiv
\Lambda[J_L]\,=\,\Theta^{-1}[J_L]\,\Phi[J_L]\,.
\end{equation}
In fact for the purpose of the calculation of the matrix element of $\hat R_a$  in eq.(\ref{defin}) the proper ordering of the factors of the charge density is equivalent to the substitution
\begin{equation}
\Omega^\dagger\,\rightarrow\, \Omega^\dagger_L\,=\,{\cal B}^\dagger_L\,{\cal C}^\dagger_L; 
\ \ \ \ \ \ \ \ \ \ \ \ \ \Omega\,\rightarrow\, \Omega_R\,=\,{\cal C}_R\,{\cal B}_R
\end{equation}
where the subscript $L$ ($R$) indicates that the respective operator depends on $J_L$ ($J_R$). 

We are now ready to proceed with the computation of the matrix element (\ref{exp}). We have
\begin{equation}\label{matex0}
\langle 0|\ \Omega^\dagger_L\, \hat R_a\,\Omega_R\ |0\rangle\ =\ \langle 0|\ \bar \Omega^\dagger_L\,\Omega_R\ |0\rangle\,,
\end{equation}
with
\begin{equation}
\bar \Omega_L\,\equiv\, \bar {\cal C}_L\,\bar {\cal B}_L;\ \ \ \ \ \ \ \ \ \ \ \ \ \ \ \ 
\bar {\cal B}_L\,\equiv\,\hat R^\dagger_a \,{\cal B}_L\,,\ \ \ \ \ \ \ \ \ \ \ \ \ \ \ \ 
\bar {\cal C}_L\,\equiv\, \hat R^\dagger_a\, {\cal C}_L\, \hat R_a\,.
\end{equation}
The action of the operator $\bar\Omega_L$ on the soft gluon vacuum is the same as the action of $\Omega_L$ with the substitution
\begin{equation}
b_{L\,i}^{a}\rightarrow \bar b_{L\,i}^{a}\,\equiv\,R^{\dagger ab}\,b_{L\,i}^{b}; 
\ \ \ \ \ \ \ \ \ \ \ \ \ \ \ \ \ \ \  \Lambda_L\rightarrow\bar\Lambda_L\,\equiv\,R^\dagger\,\Lambda_L\,R\,.
\end{equation}
Using eq.(\ref{G}) and remembering the factor $\sqrt 2$ in eq.(\ref{cohrap}) we can finally write
\begin{eqnarray}\label{matex}
\langle 0|\ \Omega^\dagger\, \hat R_a\,\Omega\ |0\rangle&=&\exp\left\{-{1\over 2}{\rm Tr}\ln (1-\bar\Lambda^\dagger\Lambda)+
{1\over 4}{\rm Tr}\ln (1-\bar\Lambda^\dagger\bar \Lambda)+
{1\over 4}{\rm Tr}\ln (1-\Lambda^\dagger \Lambda)\right\}\times\nonumber\\
&&\exp\left\{-(b-\bar b)(1-\Lambda P)[1-P\bar\Lambda^\dagger\Lambda P]^{-1}(1-P\bar\Lambda^\dagger)P(b-\bar b)\right\}\nonumber\\
&& \nonumber \\
&=&\exp\left\{-{1\over 2}{\rm Tr}\ln (1-\bar\Lambda^\dagger\Lambda)+
{1\over 4}{\rm Tr}\ln (1-\bar\Lambda^\dagger\bar \Lambda)+
{1\over 4}{\rm Tr}\ln (1-\Lambda^\dagger \Lambda)\right\}\times\nonumber\\
&&\exp\left\{-(b-\bar b)(1-P\bar\Lambda^\dagger)[1-\Lambda\bar\Lambda^\dagger]^{-1}(1-\Lambda P)P(b-\bar b)\right\}\nonumber\\
&& \nonumber \\
&=&\exp\left\{-{1\over 2}{\rm Tr}\ln (1-\bar\Lambda^\dagger\Lambda)+
{1\over 4}{\rm Tr}\ln (1-\bar\Lambda^\dagger\bar \Lambda)+
{1\over 4}{\rm Tr}\ln (1-\Lambda^\dagger \Lambda)\right\}\times\nonumber\\
&&\exp\left\{-(b-\bar b)\bar N^\dagger[\Theta\bar \Theta^\dagger-\Phi\bar \Phi^\dagger]^{-1}NP(b-\bar b)\right\}
\end{eqnarray}
where we have defined
\begin{eqnarray}
N_R&=&\Theta_R\,-\,\Phi_R\, P\,;\ \ \ \  \ \ \ \ \ \ \ \ \ \ \ \ \ \ 
\ \ \ \ \ \ \ \ \  \ \ \ \ \ \ \ \  N_L\,=\,\Theta_L\,-\,\Phi_L\, P \nonumber \\
  \bar \Theta_L&\equiv& \Theta_L\,R; \ \ \ \ \ \ \ \ \ \ \ \ \  
 \bar \Phi_L\,\equiv\,\Phi_L\,R; 
\ \ \ \ \ \ \ \ \ \ \ \ \ \ \ \bar N_L\,\equiv\, N_L\,R\,=\,\bar \Theta_L\,-\,\bar \Phi_L\, P\,.\nonumber
\end{eqnarray}
In eq.(\ref{matex}) we have omitted the subscript $L$ on barred quantities and $R$ on unbarred ones. This will be our practice also below whenever it cannot lead to confusion.

There is one point which we have glossed over, namely the ordering of the factors of the charge density entering the classical field $b$ relative to those entering $\Lambda$. It is 
obvious from our initial expressions that all factors of $\Lambda$ in eq.(\ref{matex}) have to be to the left of $b$, while all factors of $\bar \Lambda$ to the right of $\bar b$; even though we did not indicate this explicitly in eq.(\ref{matex}). As we will discuss shortly, however, this ordering is only important to subleading order in $\alpha_s$ \cite{foam}, 
and so we will not discuss it any further in the present paper.

Eq.(\ref{matex}) is far from being the end of the road. We now have to 
find $\Lambda_R$, etc explicitly in terms of the matrix $U_R$ (or equivalently in terms of the current $J_R$), and also expand the matrix element eq.(\ref{matex}) to first order in the length of the rapidity interval $Y$. We will now address the first question. 

\subsection{The matrix $\Lambda$}
To calculate the matrix $\Lambda$ consider expansion of the field $A_i^a(x^-,x)$ in terms of two different sets of creation and annihilation operators $a^\dagger, a$ and $\beta^\dagger,\beta$.

Referring to eqs.(\ref{transformed},\ref{canonical}) we can write expansion of $A$ in terms of the operators 
\begin{equation}
\beta^a_i(\eta,k)\,=\,{\cal B}\, a^a_i(\eta,k)\,{\cal B}^\dagger\,.
\end{equation}
This has the form
\begin{equation}
A^a_i(x^-,x)={1\over \sqrt 2}\int_{-\infty}^\infty {d\eta\over 2\pi}{d^2k\over 4\pi^2}
\left\{g^{(b,j,\eta,k)}(a,i,x^-,x)\,\beta^b_j(\eta,k)+g^{*(b,j,\eta,k)}(a,i,x^-,x)\,\beta^{b\dagger}_j(\eta,k)\right\}
\end{equation}
with\
\begin{eqnarray}\label{gs}
g_R\,\equiv\,
g^{(b,j,\eta,k)}(a,i,x^-,x)&=&\langle x|\ \theta(-x^-)\,e^{\,i{\partial^2\over 2k^-}\,x^-}\,(1-2l)_{ij}\,\delta^{ab}\ + \nonumber \\
&+&\theta(x^-)\,[e^{\,i{D_R^2\over 2k^-}\,x^-}\,(1-2L_R)]^{ab}_{ij}\,+\,\epsilon(x^-)\,[\Delta_R (1-l-L_R)]^{ab}_{ij}\ |k\rangle
\end{eqnarray}
where as usual $|k\rangle$ is the eigenstate of transverse momentum and $|x\rangle$ is the eigenstate of transverse coordinate. The subscript $R$ in the above expressions indicates that the relevant quantity depends on $J_R$. Here
 $L^{ab}_{ijR}\equiv D_i^{ae}(b_R)\left[{1\over D^2(b_R)}\right]^{ed}D_j^{db}(b_R)$.

The two sets of functions $(f,f^*)$ of eq.(\ref{fs}) and $(g,g^*)$ of eq.(\ref{gs}) constitute complete bases on the functional space.
It is convenient to define a scalar product on this space as
\begin{eqnarray}
&&\langle g|f\rangle^{[(c,n,\xi,p);(b,j,\eta,k)]} \,\equiv \, i\,\sum_{a,i}\,\int dx^-d^2x \,\partial^+ \,g^{*\,(c,n,\xi,p)}(a,i,x^-,x)\,f^{(b,j,\eta,k)}(a,i,x^-,x)\,; \nonumber \\
&&\langle g^*|f\rangle^{[(c,n,\xi,p);(b,j,\eta,k)]}\,
\equiv \,-i\,\sum_{a,i}\,\int dx^-d^2x \,\partial^+ \,g^{(c,n,\xi,p)}(a,i,x^-,x)\,f^{(b,j,\eta,k)}(a,i,x^-,x)\,.
\end{eqnarray}
With this scalar product the bases defined in eqs.(\ref{fs},\ref{gs}) are orthonormal:
\begin{equation}\label{orthonormality}
\langle f^{(b,j,\eta,k)}|f^{(c,l,\xi,p)}\rangle=\delta^{bc}\delta_{jl}\delta(\eta-\xi)\delta^2(k-p); \ \ \ \ \ \ \
\langle g^{(b,j,\eta,k)}|g^{(c,l,\xi,p)}\rangle=\delta^{bc}\delta_{jl}\delta(\eta-\xi)\delta^2(k-p)\,.
\end{equation}
The resolution of identity in this space (completeness relation) is given by
\begin{eqnarray}\label{completeness}
i\,\sum_{a,i}\,\int {d\eta\over 2\pi} {d^2 k\over 4\pi^2}\,
 \left[f^{(a,i,\eta,k)}(b,j,x^-,x)\right. & \partial^+ & f^{*(a,i,\eta,k)}(c,l,y^-,y)\ -
\\ 
-\ f^{*(a,i,\eta,k)}(b,j,x^-,x) & \partial^+ & \left. f^{(a,i,\eta,k)}(c,l,y^-,y)\right]\,=\,\delta^{bc}\,\delta_{jl}\,\delta(x^--y^-)
\,\delta^2(x-y) \nonumber
\end{eqnarray}
\and similarly for the basis $(g, g^*)$.

We can now calculate the matrices $\Theta$ and $\Phi$ defined in eq.(\ref{ML}).
\begin{equation}
\Theta\,=\,\langle g|f\rangle\,; \ \ \ \ \ \ \ \ \ \ \ \ \ \ \ \ \ \ \ \ \ \ 
\Phi\,=\,\langle g|f^*\rangle\,.
\end{equation}
Performing the integration we find explicitly (see Appendix B)
\begin{eqnarray}\label{MLN}
\Theta(p,\eta;k,\xi)&=&i\langle p\vert 2(l-L_R)+(1-l-L_R)\Delta_R^\dagger\vert k\rangle\ + \nonumber \\
&+&{i\over 1-e^{\xi-\eta}-i\epsilon}
\langle p\vert (1-2l)\vert k\rangle-i\langle p\vert (1-2L_R){1\over 1-{\partial^2\over D_R^2}e^{\xi-\eta}+i\epsilon}\vert k\rangle\nonumber 
\\ \nonumber && \\ 
\Phi(p,\eta;-k,\xi)&=&i\langle p\vert 2(l-L_R)+(1-l-L_R)\Delta_R^\dagger\vert k\rangle\ + \nonumber \\
&+& {i\over 1+e^{\xi-\eta}-i\epsilon}
\langle p\vert (1-2l)\vert k\rangle-i\langle p\vert (1-2L_R){1\over 1+{\partial^2\over D_R^2}e^{\xi-\eta}+i\epsilon}\vert k\rangle\nonumber\\
&& \nonumber \\ 
N(p,\eta;k,\xi)&=& \Theta(p,\eta;k,\xi)-\Phi(p,\eta;-k,\xi)\nonumber\\
&=&{2i\over e^{-(\xi-\eta)} -e^{\xi-\eta}-i\epsilon}
\langle p\vert (1-2l)\vert k\rangle-\langle p\vert (1-2L_R){2i\over {D_R^2\over\partial^2}e^{-(\xi-\eta)}-{\partial^2\over D_R^2}e^{\xi-\eta}+i\epsilon}\vert k\rangle \,.\nonumber  \\ & &
\end{eqnarray}
In the last terms in all three expression the ordering of the factors is such that all $\partial^2$ are to the right of all 
$D^2_R$. We have also not indicated explicitly the color and rotational indices on $\Theta$ and $\Phi$. Since the basis function $g$ in
 (\ref{gs}) depends on $J_R$, all the function $\Theta$, $\Phi$ and $N$  computed above are  ``Right'' quantities and should be 
understood as $\Theta_R$,  $\Phi_R$ and $N_R$. Substituting $J_L$ for $J_R$  generates the analogous ``Left'' quantities.

Mindful of the derivations in the previous subsection we find it convenient to define yet 
another set of creation and annihilation operator $\bar\beta$ and $\bar\beta^\dagger$ by
\begin{equation}
\bar\beta_\alpha\,\equiv\,
\bar{\cal B}\,a_\alpha\,\bar{\cal B}^\dagger\,=\,\bar \Theta_{\alpha\beta}\,a_\beta\,+\,\bar \Phi_{\alpha\beta}\,a^\dagger_\beta\,,
\ \ \ \ \ \ \ \ \ \ \ \ \bar \beta^\dagger_\alpha\,=\,\bar \Theta^*_{\alpha\beta}\,a_\beta\,+
\,\bar \Phi^*_{\alpha\beta}\,a^\dagger_\beta\,.
\label{barML}
\end{equation}
Using eq.(\ref{ML}) we find that the set $\beta$, $\beta^\dagger$ is related to $\bar\beta$, $\bar\beta^\dagger$
by
\begin{eqnarray}
\beta_\alpha&=&(\Theta\bar \Theta^\dagger-\Phi\bar \Phi^\dagger)_{\alpha\beta}\,\bar\beta_\beta+(\Phi\bar \Theta^T-\Theta\bar \Phi^T)_{\alpha\beta}\,\bar\beta^\dagger_\beta\\
\beta^\dagger_\alpha&=&(\Phi^*\bar \Theta^\dagger-\Theta^*\bar \Phi^\dagger)_{\alpha\beta}\,\bar\beta_\beta
+(\Theta^*\bar \Theta^T-\Phi^*\bar \Phi^T)_{\alpha\beta}\,\bar\beta^\dagger_\beta\nonumber\,.
\end{eqnarray}
We thus see that the basic matrix that enters the last line of eq.(\ref{matex}) $K=\Theta\bar \Theta^\dagger-\Phi\bar \Phi^\dagger$ 
is the transformation matrix between the operators $\beta_\alpha$ and $\bar \beta_\alpha$.
We can obtain it by calculating the overlap matrix between the appropriate basis functions 
\begin{eqnarray}\label{ke1}
K_{\alpha\beta}&=&(\Theta\bar \Theta^\dagger\,-\,\Phi\bar \Phi^\dagger)_{\alpha\beta}\,=\,\langle g_\alpha | \bar g_\beta\rangle\nonumber\\
E_{\alpha\beta}&=&(\Phi\bar \Theta^T\,-\,\Theta\bar \Phi^T)_{\alpha\beta}\,=\,\langle g_\alpha | \bar g_\beta^*\rangle\,.
\end{eqnarray}
Although we do not have the explicit form of $\bar g$, we can still use eq.(\ref{ke1}) to calculate the matrices $K$ and $E$ using the following argument. We do have the explicit form of the basis functions $g_L$, which are the same as $g_R$ except for the substitution $J_R\rightarrow J_L$. We can thus calculate explicitly the overlap matrices between $g_R$
 and $g_L$ (the analogs of $K$ and $E$). We then notice that the operator $\bar \beta$ is related to the operator $\beta_L$
by $\bar \beta=R^\dagger\beta_L $, and thus $\bar \Theta_L=\Theta_LR;\ \ \bar \Phi_L= \Phi_LR$.
Then referring to eqs.(\ref{ke1}) we see that to calculate $K$ and $E$ from the overlap matrices of $g_R$ and $g_L$ we must rotate all terms involving $b_L$ by the operator $R^\dagger$ on the left.
Performing this calculation explicitly we find
\begin{eqnarray}\label{K}
&&K(p,\eta;k,\xi)\,=\,i\langle p\vert 2(1-l-L_R)R^\dagger(l- L_L)+(1-l-L_R)(\bar\Delta-\Delta^\dagger_R)R^\dagger(1-l- L_L)\vert k\rangle\nonumber\\
&&+{i\over 1-e^{\xi-\eta}+i\epsilon}\langle p \vert(1-2l)R^\dagger(1-2l)\vert k\rangle-i\langle p\vert (1-2L_R){1\over 1-{D^2_R\over \bar D^2}e^{\xi-\eta}-i\epsilon}R^\dagger(1-2L_L)\vert k\rangle\nonumber\\ && \nonumber \\
&&E(p,\eta;-k,\xi)\,=\,i\langle p\vert 2(1-l-L_R)R^\dagger(l- L_L)+(1-l-L_R)(\bar\Delta -\Delta^\dagger_R)R^\dagger(1-l- L_L)\vert k\rangle
\nonumber \\
&&+{i\over 1+e^{\xi-\eta}}\langle p \vert(1-2l)R^\dagger(1-2l)\vert k\rangle-i\langle p\vert (1-2L_R){1\over 1+{D^2_R\over \bar D^2}e^{\xi-\eta}}R^\dagger(1-2L_L)\vert k\rangle
\end{eqnarray}
with
\begin{equation}
\bar D\ \equiv\  R^\dagger\, D_L\,R ; \ \ \ \ \ \ \ \ \ \ \ \ \ \ \ \bar\Delta\ =\ R^\dagger\,\Delta_L\,R\,.
\end{equation}
In eq. (\ref{K}) all $D_R$ are ordered to the left of all $D_L$.

Note that similarly to $\Theta$ and $\Phi$, the matrices $K$ and $E$ satisfy

\begin{equation}\label{identi}
KE^T-EK^T=0,\ \ \ \ \ \ \ \ \ \ \ \ KK^\dagger-EE^\dagger=1; \ \ \ \ \ \ \ K^\dagger K-E^TE^*=1; \ \ \ \ \ \ \ \ 
K^\dagger E-E^TK^*=0\,.
\end{equation}
For future convenience we also define the analog of the matrix $\Lambda$:
 \begin{equation}\label{defXi}
 \Xi\ =\ K^{-1}\,E\,.
 \end{equation}
 Using the identities eq.(\ref{ident},\ref{ident1},\ref{identi}) it is straightforward to show that
 \begin{equation}\label{kk}
 {1\over 1-\Xi\Xi^\dagger}\ =\ K^\dagger\, K\ ;\ \ \ \ \ \ \ \  \ \ \ \ \ \ \ \ {1\over 1-\Xi^\dagger\Xi}\ =\ K^T\, K^*\,.
 \end{equation}
 \section{$H_{RFT}$}
 Now that we have all the ingredients, we are ready to put them all together and to derive the  
RFT Hamiltonian.
Before continuing with this calculation we want to clarify the counting of powers of $\alpha_s$.
\subsection{Counting powers of $\alpha_s$}
 First, using eq.(\ref{kk}) we have
 \begin{equation}\label{traces}
 -{1\over 2}{\rm Tr}\ln (1-\bar\Lambda^\dagger\Lambda)+
{1\over 4}{\rm Tr}\ln (1-\bar\Lambda^\dagger\bar \Lambda)+
{1\over 4}{\rm Tr}\ln (1-\Lambda^\dagger \Lambda)\,=\, {1\over 4}{\rm Tr}\ln(1-\Xi^\dagger\Xi)\,=\,-{1\over 4}{\rm Tr}\ln (KK^\dagger)\,.
 \end{equation}
Now, recalling the definition (\ref{defin}) and 
referring to eq.(\ref{matex}) and eq.(\ref{traces}) we can write
\begin{equation}\label{hrft}
H_{RFT}\,=\,{d\over dY}\,\Bigg[(b-\bar b)\,\bar N^\dagger \,K^{-1}\,N\,P\,(b-\bar b)\,+\,
{1\over 4}\,{\rm Tr}\,\ln K\,K^\dagger\Bigg]\,|_{Y=0}\,.
\end{equation}
The two terms in this expression are not the same order in $\alpha_s$. We remind the reader that
this expression was derived starting with the evolution of the wave function of \cite{foam}. This evolution kept the leading in $\alpha_s$ term for arbitrary parametric value of $j$ and also terms which can be $O(1)$ at large $j$. The former terms in the evolution of the wave function give rise to the first term in $H_{RFT}$ eq.(\ref{hrft}), while the latter leads to the last term in eq.(\ref{hrft}). This can be seen directly from eq.(\ref{hrft}). As long as $j\ll 1/g$, the classical fields $b$ and $\bar b$,  as well as the operators $\Lambda$ and $\bar \Lambda$ can be expanded in powers of $gj$. To count powers of $\alpha_s$ we have to decide how to treat $R$. We know that when we calculate the forward scattering amplitude, after averaging over the projectile wave function, $R$ turns into $S$ - the $S$-matrix of scattering of a single projectile gluon on the target\cite{klwmij},\cite{zakopane}. Therefore if the target is dense we should count $R$ as being of order one  but 
 not parametrically close to the unit matrix. In this case we see that
\begin{equation}
(b-\bar b)^2\sim O(j^2); \ \ \ \ \ \ \ \ \ \ \ \ \ \ \  KK^\dagger=1+O(g^2j^2)\,.
\end{equation}
Thus indeed the first term in eq.(\ref{hrft}) is leading, while the second is suppressed by $\alpha_s$. 

When $j\sim O(1/g)$ the matrix $\Lambda$ can not be expanded. Still $\Lambda\sim O(1)$ while $b-\bar b\sim O(1/\alpha_s)$, and the first term is leading. 

If the target is not dense, the matrix $R$ is close to unity $R=1-\delta R$.
 Then $\bar b$ is perturbatively close to $b$; that is $(b-\bar b)\sim \delta R b$, and also $K\sim 1 +O(\delta R)$.
The first term in (\ref{hrft}) is $(\delta R b)^2$, and dominates over the second term, which is $(\delta R)^2$. 

Thus we conclude that in all parametric regimes of the projectile and target fields, the second term  in eq.(\ref{hrft}) is suppressed by $\alpha_s$ relative to the first term. In most of this paper we will therefore neglect the last term and will only consider the leading order piece
\begin{equation}\label{hrftl}
H_{RFT}\ =\ {d\over dY}\,\Bigg[(b-\bar b)\,\bar N^\dagger\, K^{-1}\,N\,P\,(b-\bar b)\Bigg]\,|_{Y=0}\,.
\end{equation}
At this point we are ready to comment to the question of ordering of the factors of $b$ and $\Lambda$ briefly mentioned in the subsection 3.2. 
Once we restrict ourselves to eq.(\ref{hrftl}), this ordering  becomes irrelevant. As we have noted above, in principle all factors of $\Lambda$ have to be ordered to the left of $b$, etc. However, changing the order of $\Lambda$ and $b$ brings a term proportional to the commutator 
\begin{equation}
[\Lambda,b]\ =\ {\delta \Lambda\over\delta (gb)}[gb,b ]\ \sim\  {\delta \Lambda\over\delta (gb)} \ O(g^2 b)\,.
\end{equation}
The additional terms this procedure generates are of the same order as the last term in eq.(\ref{hrft}) which we have neglected. Thus the ordering question is irrelevant as long as we are interested in the leading order term. 

The last term in eq.(\ref{hrft}) originates directly from the Bogoliubov part of the transformation $\Omega$. However, even though this term itself in eq.(\ref{hrft}) is suppressed by $\alpha_s$, it is important to realize that the Bogoliubov operator contributes in an essential way to $H_{RFT}$ and to other observables at high energy. For example the factors of $N$, $K^{-1}$ and $\bar N$ in eq.(\ref{hrftl}) are direct manifestations of the Bogoliubov part of the transformation $\Omega$. The presence of the Bogoliubov operator ${\cal B}$  also leads to interesting effects in other observables. For example, as shown in the companion paper \cite{Q}, it is responsible for the appearance of short range rapidity correlations in multi-gluon spectrum. For these particular correlations ${\cal B}$ gives the leading effect, as these correlations would be absent if the transformation $\Omega$ had only the classical piece of the type of ${\cal C}$. We discuss this in more depth in \cite{Q}.

\subsection{JIMWLK for the umpteenth time}
At this point to keep our finger on the pulse it is a good idea to perform a consistency check of our calculation so far by reproducing the JIMWLK equation. The JIMWLK equation is obtained by expanding eq. (\ref{hrftl}) to second order in $\delta/\delta j$.
Consider the expression
\begin{equation}
(b-\bar b)\,\bar N^\dagger \,K^{-1}\,N\,P\,(b-\bar b)\,.
\end{equation}
Since $b-\bar b$ is already first order in the derivative, in this approximation we have to take 
$\bar\Lambda=\Lambda$ and $\bar N=N$. This simplifies things dramatically, since in this case $K=1$.
Thus in this limit we only need to calculate
\begin{equation}\label{jimwlk1}
(b-\bar b)\,\int_{\eta,\xi,\zeta} \,N^\dagger(\eta,\xi)\,N(\xi,\zeta)\,P\,(b-\bar b)
\end{equation}
 where for simplicity we do not indicate the transverse coordinate dependences.
 Note that here and in the following the measure of the rapidity integration is defined as $d\eta/2\pi$, so that $\int_\eta\equiv \int {d\eta\over 2\pi}$.
 
Since $N(\eta,\xi)=N(\eta-\xi)$ (see (\ref{MLN})) 
is the function of rapidity difference only, one of the rapidity integrals in eq.(\ref{jimwlk1}) 
gives the factor of $Y$, and the RFT Hamiltonian becomes
\begin{equation}\label{rft}
H_{RFT}\,=\,{1\over 2\pi}\,(b-\bar b) \, N^\dagger_\perp\, N_\perp\, P\,(b-\bar b)
\end{equation}
with 
\begin{equation}
N_\perp\,\equiv \,\int_\eta \,N(\eta)\,.
\end{equation}
We remind the reader that $b$ is independent of rapidity.
The integral over the rapidity in both terms of $N$ in eq.(\ref{MLN}) is the same (after the appropriate shift of the integration variable)
\begin{equation}\label{principal}
\int d\eta {2\over e^\eta-e^{-\eta}-i\epsilon}\ =\ 
P\left[\int d\eta {2\over e^\eta-e^{-\eta}}\right]+2i\pi\int d\eta \delta (e^\eta-e^{-\eta})\ =\ i\,\pi\,.
\end{equation}
Thus we obtain
\begin{equation}\label{integral}
N_\perp\, = \ [1-l-L]\,.
\end{equation}
With this result, expanding $b-\bar b$ to first order in $\delta/\delta j$ in eq.(\ref{rft}) it is straightforward to see that eq.(\ref{rft}) reproduces the JIMWLK Hamiltonian in the form obtained in \cite{foam}.
Having established this, we now turn to the calculation of the Holy Grail - the complete $H_{RFT}$.

\subsection{ The Holy Grail: the RFT Hamiltonian.}
We have to evaluate 
\begin{eqnarray}
[b-\bar b]\,\bar N^\dagger\, K^{-1}\,NP\,[b-\bar b]&=&\nonumber \\
=\,\int_{x,y,z,u}\left[b(x)-\bar b(x)\right]&&\int_{\eta,\xi,\lambda\zeta} 
\bar N^\dagger(\eta,x;\xi,y)\,K^{-1}(\xi,y;\lambda, z)\,N(\lambda,z;\zeta, u)\,\left[b(u)-\bar b(u)\right]\,.
\end{eqnarray}
Note that all the operators in question ($\bar N$, $N$ and $K$) depend only on rapidity differences.
 Thus the integration over rapidities yields
\begin{equation}\label{eqs}
{1\over 2\pi}\,\int_{x,y,z,u}[b(x)-\bar b(x)]\, \bar N_\perp^\dagger(x,y)\,K_\perp^{-1}(y, z)\,N_\perp(z,u)\,[b(u)-\bar b(u)]\,Y\,.
\end{equation}
Here $Y$ is the total rapidity interval  and we have defined
\begin{equation}
K_\perp(z,u)\ \equiv\ \int {d\eta\over 2\pi}\, K(\eta,z;\xi, u)\,. 
\end{equation}
Note that since the operator $K$ depends only on rapidity difference, the integral of 
the inverse is the same as the inverse of the integral
\begin{equation}
\int_\eta K^{-1}(\eta-\xi)\ =\ \left[\int_\eta K(\eta-\xi)\right]^{-1}\,.
\end{equation}
If all rapidity integrals were finite,
the expression in eq.(\ref{eqs}) would be
 proportional to the total length of rapidity interval opened by the boost, 
or "created"  by the action of the operator $\Omega$. Eq.(\ref{defin}) then would imply
\begin{equation}
H_{RFT}\ =\ {1\over 2\pi}\,\int_{x,y,z,u}\,[b(x)-\bar b(x)]\, \bar N_\perp^\dagger(x,y)\,
K_\perp^{-1}(y, z)\,N_\perp(z,u)\,[b(u)-\bar b(u)]\,.
\end{equation}
Using eqs.(\ref{principal},\ref{integral}) we have
\begin{equation}
N_\perp(x,y)\ =\ 1-l-L_R;\ \ \ \ \ \ \ \ \ \ \ \ \ \ \ \ \ \ \bar N_\perp(x,y)\ =\ \left(1-l- L_L\right)\,R\,.
\end{equation}
The integration of $K^{-1}$ however gives a puzzling result. If expression eq.(\ref{K}) is taken literally, we find that at large $Y$
\begin{equation}\label{intk}
\int_\eta K(\eta,\xi)\propto Y; \ \ \ \ \ \ \ \ \ \ \ \ \ \ \ \ \ \ \int_\eta K^{-1}(\eta,\xi)\propto {1\over Y}
\end{equation}
It then seems that eq.(\ref{eqs}) does not have the overall factor $Y$ and thus does not yield a logarithmic evolution. The resolution of this puzzle is the following. Since $H_{RFT}$ is obtained by differentiation with respect to $Y$, at this stage of the calculation we really should consider $Y$ to be small rather than large.  In writing eq.(\ref{eqs}) we have assumed that it does not contain higher powers of $Y$. On the other hand our expressions when expanded in powers of $\alpha_s$ at small $j$ contain higher order perturbative contributions. The standard perturbative expansion at any fixed order (beyond the leading one) contains higher powers of $Y$. The simplest source of higher powers
of rapidity is the "iteration" of  the leading order kernel. Such terms have been identified
and removed in the NLO computation of the JIMWLK kernel  \cite{nexttoleading}.
Obviously in our case also these higher powers of logarithms have to be subtracted in order to extract the evolution kernel. This of course is the same as treating $Y$ as infinitesimal while extracting $H_{RFT}$ from the fixed order perturbative calculation. Clearly in the present case we should do the same - we should learn how to subtract the higher powers of $Y$ in eq.(\ref{eqs}). That such higher powers exist is quite obvious. If we  think of $Y$ as of finite range of integration of the rapidity integrals, eq.(\ref{intk}) reads schematically as
\begin{equation}\label{intk1}
\int_\eta K(\eta,\xi)\,=\,
\kappa_1\,+\,\kappa_2\,Y;\ \ \ \ \ \ \ \ \ \ \  \ \ \ \ \ \ \ \ \ \ \ \ \ 
\int_\eta K^{-1}(\eta,\xi)\,=\,\kappa_1^{-1}\,\sum_{n=0}^\infty(-1)^n(\kappa_1^{-1}\,\kappa_2\,Y)^n
\end{equation}
with $\kappa_1$ and $\kappa_2$ some operators in the transverse space. 
This structure is demonstrated explicitly in the Appendix C, where we perform the perturbative expansion of eq.(\ref{eqs}). We show there that the next to leading term contains a term proportional to $Y^2$, as in \cite{nexttoleading}. 

In view of this it is clear that our task is to extract the $n=0$ term in the sum in eq.(\ref{intk1}). This is actually not difficult. Examining the expression eq.(\ref{K}) it becomes clear that the  $\kappa_2\,Y$ term in eq.(\ref{intk1})  comes from the integral over rapidity of the rapidity independent terms in $K$. Let us rewrite $K$ as follows
\begin{equation}
K(\eta-\xi)\,=\,\tilde K(\eta-\xi)\,+\,K_A(\eta-\xi)\,+\, K_C
\end{equation}
where $\tilde K$ is a symmetric ($\tilde K(\eta-\xi)=\tilde K(\xi-\eta)$), bounded operator whose matrix elements vanish as $|\eta-\xi|\rightarrow\infty$; 
while $K_A$ is antisymmetric ($K_A(\eta-\xi)=-K_A(\xi-\eta)$) and $K_C$ does not depend on rapidity. 
We can then write
\begin{equation}
\int_\eta K(\eta-\xi)\ =\ \int_\eta\tilde K(\eta-\xi)\,+\,K_C\,Y\,.
\end{equation}
Thus expansion of $\int_\eta K^{-1}$ in powers of $Y$ is the same as expansion in powers of $K_C$. Since we are interested only in the leading term in this expansion, we need to drop $K_A+K_C$ from the expression for $K$ in eq.(\ref{K}) and retain $\tilde K$.  Eq.(\ref{hrftl}) now reads
\begin{equation}\label{rft1}
H_{RFT}\ =\ {1\over 2\pi}\,
\int_{x,y,z,u}[b(x)-\bar b(x)] \,\bar N_\perp^\dagger(x,y)\,\tilde K_\perp^{-1}(y, z)\,N_\perp(z,u)\,[b(u)-\bar b(u)]\,.
\end{equation}
A quick calculation gives
\begin{eqnarray}
\tilde K(\eta,\xi)&=&{i\over 2}\,(1-2l)\,R^\dagger\,(1-2l)\,
\left[ {1\over 1-e^{\xi-\eta}+i\epsilon}+{1\over 1-e^{-(\xi-\eta)}+i\epsilon}-1\right]\nonumber\\
&-&{i\over 2}(1-2L_R)\left[{1\over 1-{D^2_R\over \bar D^2}e^{\xi-\eta}-i\epsilon}+{1\over 1-{D^2_R\over \bar D^2}e^{-(\xi-\eta)}-i\epsilon}-1\right]R^\dagger(1-2L_L)
\end{eqnarray}
To calculate the integral over rapidity, consider an integral
\begin{equation}
\int dx \left[{1\over 1-ae^x-i\epsilon}+{1\over 1-ae^{-x}-i\epsilon}-1\right]\ =\ 
\int dx {1-a^2\over (1-ae^x-i\epsilon)(1-ae^{-x}-i\epsilon)}\,.
\end{equation}
To integrate we can close the contour in the complex plane. The integrand decays exponentially at infinity everywhere except close to the imaginary axis. Nevertheless the contour can be closed. To see this note that the integrand has an infinite number of poles at $x^-_n=-\ln a+2\pi in-i\epsilon$ and $x^+_n=\ln a+2\pi in+i\epsilon$. The residue of the pole $x^-_n$ is the negative of the residue of the pole $x^+_n$. Thus the contribution of a pair of poles vanishes, and the result of the integral does not depend on how many pairs of poles the contour encloses. Closing the contour above the real axis we see that the only uncanceled pole is $x^+_0$, it's residue is unity, and the result of the integration is
\begin{equation}
\int dx \left[{1\over 1-ae^x-i\epsilon}+{1\over 1-ae^{-x}-i\epsilon}-1\right]\ =\  2\pi i\,.
\end{equation}
Analogously
\begin{equation}
\int dx \left[{1\over 1-ae^x+i\epsilon}+{1\over 1-ae^{-x}+i\epsilon}-1\right]\ =\ -2\pi i\,.
\end{equation}
Putting it all together we find
\begin{equation}
\tilde K_\perp\,=\,{1\over 2}\,\left[(1-2l)\,R^\dagger\,(1-2l)\,+\,(1-2L_R)\,R^\dagger\,(1-2L_L)\right]\,.
\end{equation}
Thus the complete RFT Hamiltonian is given by
\begin{eqnarray}\label{hg}
H_{RFT}& =&
{1\over \pi}\,[b_RR^\dagger- b_L\,]  \,\left(1-l- L_L\right) \nonumber \\ && \times\  
\left[(1-2l)\,R^\dagger\,(1-2l)\,+\,(1-2L_R)\,R^\dagger\,(1-2L_L)\right]^{-1}\ \left(1-l- L_R\right)\ [b_R-R^\dagger \,b_L]\,.
\end{eqnarray}
This is the main result of the present paper. Note, that $H_{RFT}$ can be written entirely in terms of three unitary matrices $R(x),\ U_R(x)$ and $U_L(x)$. Restoring all the indices and transverse coordinate dependences we have
\begin{eqnarray}\label{hg1}
H_{RFT}&=&
{1\over 8\pi^3 }\,\int_{x,y,z,\bar z}[b_{Ri}^b(x)\,R^{\dagger ba}(x)\,-\,b^a_{Li}(x)]\ \left[\delta_{ij}{1\over (x-z)^2}\,-\,
2\,{(x-z)_i\,(x-z)_j\over (x-z)^4}\right]\nonumber \\ \nonumber \\ 
&\times&\left[\delta^{ac}\,+\, [U_L^\dagger(x)\,U_L(z)]^{ac}\right]
\  \tilde K^{-1\,cd}_{\perp jk}(z,\bar z)\ \left[\delta_{kl}{1\over (y-\bar z)^2}\,-\,2\,{(y-\bar z)_k\,(y-\bar z)_l \over (y-\bar z)^4}\right]
\nonumber \\ \nonumber \\ 
&\times&\left[\delta^{de}\,+\, [U_R^\dagger(\bar z)\,U_R(y)]^{de}\right]\ [b_{Rl}^e(y)\,-\,R^{\dagger ef}(y)\,b^f_{Lk}(y)]
\end{eqnarray}
with
\begin{eqnarray}
\tilde K^{\ ab}_{\perp ij}(x,y)&=&{1\over 2\,\pi^2}\,\int_z\left[\delta_{ik}{1\over (x- z)^2}\,-\,2\,{(x- z)_i\,(x- z)_k\over (x- z)^4}\right]
\ \left[\delta_{kj}{1\over (z-y)^2}\,-\,2\,{(z-y)_k\,(z-y)_j\over (z-y)^4}\right]\nonumber\\
&\times&\left\{R^{\dagger ab}(z)\,+\,\left[U_R^\dagger(x)\,U_R(z)\,R^{\dagger }(z)\,U^\dagger_L(z)\,U_L(y)\right]^{ab}\right\}
\end{eqnarray}
and
\begin{equation}
b^a_{L(R)i}\,=\,-{1\over g}\,f^{abc}U_{L(R)}^{\dagger bd}\,\partial_i \,U_{L(R)}^{dc}\,.
\end{equation}

We note that in appropriate limits eq.(\ref{hg}) reduces to the known results. We have already discussed the JIMWLK limit, and it also  directly follows from the above formula. The leading order expansion of $R$ in powers of $\delta/\delta j$ requires that we set $R=1$ everywhere except in $b_LR$, which has to be expanded to first order. In this limit eq.(\ref{hg}) reduces to $H_{JIMWLK}$. In the limit of the low charged density on the other hand, we have to expand in $j$.
To lowest order in $j$ we have $L_R=L_L=l$. The denominator in eq.(\ref{hg}) cancels the two adjacent factors $(1-2l)$ in the numerator leaving the factor $1/2$. Additionally we have to expand $b$ to first order in $j$, that is $b_i={\partial_i\over\partial^2}j$. We thus obtain the KLWMIJ limit \cite{klwmij}
\begin{equation}\label{Hklwmij}
H_{RFT}\rightarrow H_{KLWMIJ}\,=\,{1\over 2\pi}\,
\left[j\,{\partial_i\over\partial^2}\,-\,j\,R^\dagger{\partial_i\over\partial^2}\,R\right]\, 
\left[{\partial_i\over\partial^2}\,j\,-\,R^\dagger\,{\partial_i\over\partial^2}\,R\,j\right]\,.
\end{equation}

Note that if we do not expand $b$ to first order on $j$ but still set $L_R=L_L=l$ in the rest of the expression, we obtain the form dubbed in \cite{klwmij} "KLWMIJ+"
\begin{equation}\label{klwmij+}
H_{KLWMIJ+}\ =\ {1\over 2\pi}\,[b_R-R^\dagger b_L]^2\,.
\end{equation}
Although this looks simple and rather appealing, it is not a leading term in $H_{RFT}$ in any well defined limit. 
This expression was also derived in \cite{SMITH,Balitsky05}. It has been suggested in \cite{SMITH} 
that it might in fact be the complete expression for $H_{RFT}$, including the Pomeron loop effects. Our present derivation makes it clear that this is not the case, as in general the terms kept in $H_{KLWMIJ+}$ are as important as the terms omitted.  In particular none of the effects coming from the Bogoliubov part of the operator $\Omega$ is taken into account in eq.(\ref{klwmij+}) and thus the JIMWLK limit is not reproduced\footnote{Ref.\cite{SMITH} asserts that eq.(\ref{klwmij+}) does reproduce the JIMWLK limit. The argument of \cite{SMITH} however is not direct but is rather based on an application of the Dense-Dilute Duality transformation \cite{something} $U_R\rightarrow R$. To be able to apply this transformation to $H_{RFT}$ however one has to know the transformation properties of $U_L$. A certain transformation of $U_L$ was postulated in \cite{SMITH}. Unfortunately this did not take into account the fact that $U_L$ is not independent of $U_R$ and $R$, but rather is 
 unambiguously determined once $U_R$ and $R$ are known. The transformation postulated in \cite{SMITH} turns out to be inconsistent with this dependence, and thus the argument about reproducing the JIMWLK limit given in \cite{SMITH} fails.}.

\section{Discussion.}
First, let us summarize the main results of this paper. We have calculated the Hamiltonian of the Reggeon field theory in the eikonal approximation. 
\begin{eqnarray}\label{f1}
&&H_{RFT}\ =\ {1\over \pi}\,[b_R- b_L\,R]\, R^\dagger\, \left(1-l- L_L\right)\\
&&\ \ \ \ \ \ \ \ \ \ \ \  \ \ \ \ \ \ \ \ \ \ \ \ \times\ 
\left[(1-2l)\,R^\dagger\,(1-2l)\,+\,(1-2L_R)\,R^\dagger\,(1-2L_L)\right]^{-1}\ 
\left(1-l- L_R\right)\,[b_R-R^\dagger \,b_L]\,.\nonumber
\end{eqnarray}
This Hamiltonian governs the evolution of hadronic observables with energy. It reproduces the JIMWLK and KLWMIJ limits discussed in the literature earlier, as well as the "intermediate" form - KLWMIJ+. The evolution generated by this Hamiltonian is valid for all interesting values of the color charge density $g<j<{1\over g}$. This covers the hadronic targets from a dilute perturbative object (a "dipole") to a dense non-perturbative object with large gluon density (a "nucleus"). Importantly it also describes the evolution of a nuclear projectile at all momentum scales. This is important since at high transverse momentum - larger than the saturation momentum $Q_s$, a nucleus is also transparent, and thus the color charge density at these momenta is small. Thus if one is indeed interested to follow evolution at large momentum scales, the JIMWLK equation is not adequate even if the projectile is a nucleus. Another regime where JIMWLK is not adequate for a nuclear target is in description of nuclear periphery where the color charge density is also small. Again, in this regime $H_{RFT}$ supercedes $H_{JIMWLK}$, although here the situation is complicated by non-perturbative soft effects related to confinement physics.

We note that the Hamiltonian eq.(\ref{f1}) is invariant under the 
Gribov`s signature symmetry, which in \cite{seven} was identified as the symmetry under the transformation
$R\rightarrow R^\dagger,\ \ \ J_L\rightarrow J_R$.

An attempt to derive the complete Reggeon Field Theory Hamiltonian was made in \cite{SMITH}. As we mentioned earlier, the result presented in \cite{SMITH} is the KLWMIJ+ Hamiltonian, while the corrections due to the Bogoliubov part of the operator $\Omega$ were missing. Since the derivation of \cite{SMITH} is in the Lagrangian formalism, it also leaves unanswered the question about the canonical structure of the operators entering the RFT Hamiltonian. Our derivation clarifies this question along the lines discussed in \cite{yinyang}. The Hilbert space of the Reggeon field theory is the space of the functionals of the unitary matrix $R$:  $\Psi[R(x)]$. The unitary matrix $R$ is therefore the only independent quantum degree of freedom in this theory. The operators $J_L(x)$ and $J_R(x)$ act on this space as the generators of local left and right $SU(N)$ transformation group. Consequently the operators $U_L$ and $U_R$ do not commute with $R$, but  have rather complicated commutation relations determined by the dependence of $U_R$ on $J_R$ and of $U_L$ on $J_L$. On the other hand $U_R$ commutes with $U_L$, since $J_R$ and $J_L$ commute with each other.

There is a question, which we have not touched upon so far, but which is important for understanding the consistency of the whole approach. The issue is when can we consistently treat the length of the rapidity interval $Y$ as large, and when can we treat it as small. In our derivation we do both.
First, when deriving the evolution of the hadronic wave function in \cite{foam}, we have approximated the interaction between the soft modes and the valence modes of the gluon field by the eikonal vertex\footnote{ This is technically not the same as the eikonal approximation for scattering on the target, although the consistency of the whole approach likely requires both.}. This is valid as long as the main contribution to physical quantities comes from the soft gluons with rapidities much smaller than the rapidities of the valence ones. The borderline region, where the rapidities of the "soft" and "valence" gluons are comparable, is of order $\delta Y\sim 1$. Thus consistency of the eikonal approximation requires $Y\gg 1$. As long as the eikonal approximation is valid, we can ignore finiteness of $Y$ and treat the phase space available to soft gluons as infinite, which is what was done in \cite{foam}.
 
On the other hand when calculating $H_{RFT}$ itself we have differentiated the matrix element with respect to $Y$ at $Y=0$. Thus at this point we have treated $Y$ as small.
Even though it may seem odd, in principle the two approximations are not necessarily incompatible. The question is how fast is the evolution of a given physical quantity. For example, in the BFKL calculation
to get significant change in the cross section due to the contribution of soft gluons, one needs to increase the rapidity of a process by $Y\sim {1\over \alpha_s}$. This means, that when taking the derivative to derive the BFKL Hamiltonian the smallness of the step in $Y$ means simply $\Delta Y\ll {1\over \alpha_s}$. At small coupling constant this is still compatible with the condition for validity of the eikonal approximation $\Delta Y\gg 1$. We can thus use the eikonal approximation ("large" $Y$ approximation) to calculate the evolution even on a "small" rapidity interval. One can have one's cake and eat it! 

The same exact argument holds for the KLWMIJ and JIMWLK Hamiltonians. Formally the evolution eigenvalues in both cases are of order $\alpha_s$. To see this remember that rescaling $j$, the KLWMIJ Hamiltonian can be written as
\begin{equation}
H_{KLWMIJ}\,=\alpha_s \,\bar H_{KLWMIJ}[{1\over g}j,R]\,.
\end{equation}
We use this Hamiltonian when $j\sim g$ and $R\sim 1$, thus parametrically the eigenvalues are $
\omega_{KLWMIJ}\sim O(\alpha_s)$.
Likewise for JIMWLK
\begin{equation} H_{JIMWLK}\,= \alpha_s \,\bar H_{JIMWLK}[U,{\delta\over \delta (gj)}]
\end{equation}
where $U\sim 1$ and  $gj\sim 1$ so that $\omega_{JIMWLK}\sim O(\alpha_s)$\footnote{In fact we know that due to the dense-dilute symmetry \cite{something} the two sets of eigenvalues are equal.}. Thus for the evolution of the forward scattering amplitude in either KLWMIJ or JIMWLK limits, we have
\begin{equation}
S(Y)\,\sim\, e^{\alpha_s \,\kappa\, Y}\,S(0)
\end{equation}
where $\kappa$ is of order one. Thus again we see that the $S$-matrix changes significantly only if the rapidity is large $Y\sim {1\over \alpha_s}$.
So in the KLWMIJ and JIMWLK limits one can have large rapidity interval to accommodate the eikonal approximation, but still small enough steps in rapidity to be able to calculate the derivative.

This argument cannot be straightforwardly extended for $H_{RFT}$ of eq.(\ref{f1}). Assuming $b-\bar b\sim O(1/g)$ and $R\sim 1$ we expect $H_{RFT}\sim O(1/\alpha_s)$, and thus $\omega_{RFT}\sim {1\over \alpha_s}$. This suggests that the evolution with $H_{RFT}$ is very fast and observables change significantly over the rapidity range $Y\sim O(\alpha_s)$. If this were indeed the case, the whole approach beyond the KLWMIJ/JIMWLK limits would be in question. However, the situation is not really so bad. Although we do not know the spectrum of $H_{RFT}$ it is very likely that just like in the JIMWLK/KLWMIJ case the spectrum is continuous with lowest eigenvalue being zero. In fact the existence of eigenstates with zero eigenvalue is straightforward to establish. Just like for JIMWLK/KLWMIJ the state (in the RFT Hilbert space) with zero charge density $j\vert {\rm Yang} \rangle=0$  and the state with wave function independent of $j$: $R\vert {\rm Yin}\rangle=\vert {\rm Yin}\rangle$
  are both eigenstates of $H_{RFT}$ with zero eigenvalues \cite{yinyang}. Physically those are the totally white (vacuum) and the totally black (black disk) states.
If indeed the spectrum of $H_{RFT}$ is continuous then the spectrum also has eigenvalues of $O(\alpha_s)$. Evolution of observables which have significant overlap with eigenstates that correspond to these eigenvalues then has the same speed as for the JIMWLK/KLWMIJ evolution. More formally we can represent any observable and its evolution as
\begin{equation}
\langle O\rangle\ =\ \int [dj]\,W^P[j]\ O[j]\ =\ \langle W^P\vert O\rangle
\end{equation}
with $O[j]$ is itself a target average of the original observable operator
\begin{equation}\label{aver}
O[j]=\int [dS] O[j,S]W^T[S]\,.
\end{equation}
and bra and ket referring to states in the Hilbert space of RFT.
Then
\begin{equation}
{d\over dY} \langle O\rangle=-\int [dj]W^P[j]H_{RFT}\,O[j]=-\sum_n\omega_n\langle W^P\vert n\rangle\langle n\vert O\rangle; \ \ \ \ \ \langle O\rangle_Y=\sum_n e^{-\omega_nY}\langle W^P\vert n\rangle\langle n\vert O\rangle\,.
\end{equation}
Thus if at the initial rapidity both $\vert W^P\rangle$ and $\vert O\rangle$ have overlap with eigenstates $n$ such that $\omega_n\sim O(\alpha_s)$, the evolution of such observables with $H_{RFT}$ is slow. In the companion paper \cite{Q} we discuss an argument purporting to establish
that  inclusive multi-gluon amplitudes evolve with the JIMWLK Hamiltonian. If true, this would then suggest that the multi-gluon amplitudes are indeed observables of this sort.  

An interesting question is whether the forward scattering amplitude in the nucleus-nucleus collision is also such an observable. As we have discussed above, the forward amplitude is not evolved with JIMWLK Hamiltonian, but with complete $H_{RFT}$. Also if we consider scattering on a fixed configuration of the target field $\alpha$, the evolution is certainly fast. The counting which gives $H_{RFT}$ as $O(1/\alpha_s)$ clearly holds for a generic configuration of $\alpha\sim 1/g$. A small change of the projectile charge density $\delta j\sim O(1)$ leads to a large change of the phase since $\alpha$ is very large. Another way of putting it, is that the observable $\exp\{ij\alpha\}$ is orthogonal to $\vert {\rm Yin}\rangle$ and probably also almost orthogonal to other Yin -like states, while $W^P[j]$ being the weight function of a nucleus, has a very small overlap with $\vert {\rm Yang}\rangle$ and Yang -like states. Thus the overlap of $W^P$ with $\exp\{ij\alpha\}$ is dominated
  by the states $n$ with large eigenvalues $\omega_n$. 

The interesting quantity is however not the scattering amplitude on a fixed configuration of the field $\alpha$, but rather the target-averaged observable like in eq.(\ref{aver}). Averaging over the target fields has the effect of averaging the phase $\exp\{ij\alpha\}$ to an extremely small value. Thus it could well be that the contribution of the states with large $\omega_n$ is erased by the averaging procedure. If this is the case we are back to the situation where the states with small $\omega_n$ dominate both the value of the observable and its evolution. In this case the eikonally derived $H_{RFT}$ is applicable also for the forward scattering amplitude. 
We have not studied this question any further and at the moment cannot make a definitive statement. 

Finally we want to mention several possible lines for further research. 
There are two questions that can be addressed immediately. 
First is extending our calculation \cite{Q} of multi-gluon spectrum to the situation when the rapidities of the observed gluons are far from each other in rapidity. This should follow very closely the derivation of \cite{multig} and should not pose any major complications. 

Secondly, we have not studied here the effects of the ${\rm Tr} \ln$ term in $H_{RFT}$, since it is a subleading in the coupling constant effect. At $j<1/g$ this term collects some of the subleading corrections, but may not be complete (we have not analyzed this question). However, as follows from \cite{foam}, at $j\sim 1/g$ this term contains all the NLO corrections to $H_{JIMWLK}$. Thus for large projectile fields it is very interesting to calculate this term explicitly, as it is the generalization of purely gluonic terms in the calculation of \cite{nexttoleading} for the case when the projectile is dense (and the target, in principle is arbitrary). 

A more formal question is that about the selfduality of $H_{RFT}$. It was shown in \cite{something} that the RFT Hamiltonian must be selfdual under the Dense Dilute Duality transformation $U_R\leftrightarrow R$. We have not addressed this question in the present paper. The selfduality is not an entirely trivial issue, since to verify it we need to understand the duality transformation properties of $U_L$. We are hopeful that this can be done and plan to address this question in the future.

Finally, of course it is imperative to understand what physical consequences has the inclusion of Pomeron loops in the evolution. We hope that it will be possible to analyze the dynamics generated by $H_{RFT}$ at least numerically. It appears to be a much more complicated problem than, for example JIMWLK evolution successfully studied in the Langevin formulation in \cite{langevin}, as the kernel now has an infinite number of derivatives and does not allow Langevin formulation. Still we are hopeful that some quantitative analysis will be possible. It is generally true, that it is one matter to find the Holy Grail, and quite another to learn to drink from it.


 \section*{Acknowledgments}

TA, AK and JP acknowledge support from the DOE through the grant DE-FG02-92ER40716. The work 
of ML is partially supported by the DOE grants DE-FG02-88ER40388 and DE-FG03-97ER4014.

\appendix

\section{Appendix -
The calculation of matrix elements.} \label{sec:A}

In this appendix we give the details of the calculations in Section 3.
We  have to calculate the matrix element
\begin{equation}
F[\Lambda,\bar \Lambda]\,=\,\langle 0|e^{-{1\over 2}\,a\,\bar\Lambda^\dagger \,a}\,
e^{-{1\over 2}\,a^\dagger\,\Lambda \, a^\dagger}|0\rangle\,.
\end{equation}
Differentiating this with respect to $\Lambda$ we obtain
\begin{eqnarray}
{\partial F\over \partial \Lambda_{\alpha\beta}}&=&\langle 0\vert
e^{-{1\over 2}\,a\,\bar \Lambda \,a}\,\left[-{1\over 2} \,a^\dagger_\alpha\, a^\dagger_\beta\right]\,
e^{-{1\over 2} a^\dagger\,\Lambda \,a^\dagger}\vert 0\rangle\nonumber\\
&=&\langle 0\vert
{\partial\over \partial a_\alpha}{\partial\over \partial a_\beta}\,e^{-{1\over 2}\,a\,\bar \Lambda \,a}\,
e^{-{1\over 2}\,a^\dagger\, \Lambda \, a^\dagger}\vert 0\rangle\nonumber\\
&=&\langle 0\vert
e^{-{1\over 2}\,a\,\bar \Lambda \,a}\,
[-\bar \Lambda_{\alpha\beta}+\bar\Lambda_{\alpha\gamma}\bar\Lambda_{\beta\delta} a_\gamma a_\delta]\,
e^{-{1\over 2}\,a^\dagger\, \Lambda \, a^\dagger}\vert 0\rangle\nonumber\\
&=&{1\over 2}\bar\Lambda^\dagger_{\alpha\beta}F+[\bar\Lambda^\dagger {\partial F\over \partial \bar \Lambda^\dagger}\bar\Lambda^\dagger]_{\alpha\beta}\,.
\end{eqnarray}
The resulting differential equation:
\begin{equation}
{\partial F\over \partial \Lambda_{\alpha\beta}}={1\over 2}\bar\Lambda^\dagger_{\alpha\beta}F+[\bar\Lambda^\dagger {\partial F\over \partial \bar \Lambda^\dagger}\bar\Lambda^\dagger]_{\alpha\beta}
\end{equation}
is solved by
\begin{equation}
F[\Lambda,\bar\Lambda]\ =\ \exp\left\{-{1\over 2}{\rm Tr} \ln(1-\bar\Lambda^\dagger\Lambda)\right\}\,.
\end{equation}

We now turn to the matrix element
\begin{equation}
G[\gamma_\alpha]\,=\,
\langle 0|e^{-{1\over 2}\,a\,\bar\Lambda^\dagger\, a}\,e^{-i\bar b\,(a+P\,a^\dagger)}
\,e^{-i b \,(a+P\,a^\dagger)}\,e^{-{1\over 2}\,a^\dagger\,\Lambda \, a^\dagger}|0\rangle=\langle 0|e^{-{1\over 2}\,a\,
\bar\Lambda^\dagger a}\,
e^{-i \gamma\,(a+P\,a^\dagger)}\,e^{-{1\over 2}\,a^\dagger\,\Lambda \, a^\dagger}|0\rangle
\end{equation}
where we have introduced notation
\begin{equation}
\gamma_\alpha=b_\alpha-\bar b_\alpha\,.
\end{equation}
To calculate this matrix element we differentiate with respect to $\gamma_\alpha$. 
\begin{eqnarray}\label{a1}
{\partial G\over\partial \gamma_\alpha}&=&i\langle 0|e^{-{1\over 2}\,a\,\bar\Lambda^\dagger \,a}\,
(a_\alpha+P_{\alpha\beta}a^\dagger_\beta)\,
e^{-i \gamma(a+Pa^\dagger)}\,e^{-{1\over 2}\,a^\dagger\,\Lambda\,a^\dagger}|0\rangle\nonumber\\
&=&i\langle 0|\Big[P_{\alpha\beta}{\partial\over \partial a_\beta}\,
\Big[e^{-{1\over 2}\bar\Lambda^\dagger aa}\Big]+ e^{-{1\over 2}\bar\Lambda^\dagger aa}a_\alpha\Big]
e^{-i \gamma(a+Pa^\dagger)}e^{-{1\over 2}\Lambda a^\dagger a^\dagger}|0\rangle\nonumber\\
&=&i\Big(\delta_{\alpha\beta}-(P\bar \Lambda^\dagger)_{\alpha\beta}\Big)\langle 0| e^{-{1\over 2}\,a\,\bar\Lambda^\dagger \,a}\,
a_\beta\,e^{-i \gamma(a+Pa^\dagger)}\,e^{-{1\over 2}\,a^\dagger\,\Lambda\, a^\dagger}|0\rangle\nonumber\\
&=&i\Big(\delta_{\alpha\beta}-(P\bar \Lambda^\dagger)_{\alpha\beta}\Big)\,
\left[i(\gamma P)_\beta G+\langle 0| e^{-{1\over 2}\,a\,\bar\Lambda^\dagger \,a}\,
e^{-i \gamma(a+Pa^\dagger)}\,(-\Lambda_{\beta\gamma}a^\dagger_\gamma)\,e^{-{1\over 2}\,a^\dagger\,\Lambda \, a^\dagger}|0\rangle
\right]\,.
\end{eqnarray}
The last term can be expressed in terms of ${\partial G\over \partial\gamma_\alpha}$ if in the very first line instead of commuting $a^\dagger$ to the left, we commute $a$ to the right
\begin{eqnarray}
{\partial G\over\partial \gamma_\alpha}&=&i\langle 0|e^{-{1\over 2}\,a\,\bar\Lambda^\dagger \,a}
\,(a_\alpha+P_{\alpha\beta}a^\dagger_\beta)\,
e^{-i \gamma(a+Pa^\dagger)}e^{-{1\over 2}\,a^\dagger\,\Lambda \, a^\dagger}|0\rangle\nonumber\\
&=&i\Big(\delta_{\alpha\beta}-(\Lambda P)_{\alpha\beta}\Big)\langle 0| e^{-{1\over 2}\,a\,\bar\Lambda^\dagger \,a}
e^{-i \gamma(a+Pa^\dagger)}(Pa^\dagger)_\beta \,e^{-{1\over 2}\,a^\dagger\,\Lambda\, a^\dagger}|0\rangle
\end{eqnarray}
So that 
\begin{equation}
\langle 0| e^{-{1\over 2}\,a\,\bar\Lambda^\dagger \,a}
e^{-i \gamma(a+Pa^\dagger)}a^\dagger_\alpha e^{-{1\over 2}\,a^\dagger \,\Lambda \, a^\dagger}|0\rangle=-i\Big[ P{1\over 1-\Lambda P}\Big]_{\alpha\beta}{\partial G\over\partial\gamma_\beta}\,.
\end{equation}
Together with eq.(\ref{a1}) this gives equation
\begin{equation}\label{G}
{\partial G\over\partial \gamma_\alpha}=-(1-P\bar\Lambda^\dagger)_{\alpha\beta}P_{\beta\omega}\gamma_{\omega} G-(1-P\bar\Lambda^\dagger)_{\alpha\beta}[\Lambda P]_{\beta\epsilon}(1-\Lambda P)_{\epsilon\omega}^{-1}{\partial G\over\partial \gamma_\omega}\,.
\end{equation}
This is solved by
\begin{eqnarray}
G[\gamma]&=&G[0]\exp\{-{1\over 2}\gamma(1-\Lambda P)[1-P\bar\Lambda^\dagger\Lambda P]^{-1}(1-P\bar\Lambda^\dagger)P\gamma\}\\
&=&G[0]\exp\{-{1\over 2}\gamma (1-P\bar\Lambda^\dagger )[1-\Lambda\bar\Lambda^\dagger ]^{-1}(1-\Lambda P)P\gamma\}\nonumber
\end{eqnarray}
where $G[0]=F[\Lambda,\bar\Lambda]$ of eq.(\ref{F}).

\section{Appendix - The overlap matrices.}
\label{sec:B}
In this Appendix we calculate the overlap matrices $\Theta$ and $\Phi$ used in the text. Here we will use frequency $p^-$ rather than rapidity to label the basis functions. However the normalization is still taken to be consistent with expanding the field $A$ in terms of creation and annihilation operators $a_\eta,\ \ a^\dagger_\eta$. We start from the expression for the basis functions 
\begin{eqnarray}
&&f^{(b,j,\eta,k)}(a,i,x^-,x)\,=\,\delta_{ij}^{ab}\langle x\vert e^{i{\partial^2\over 2 k^-}x^-}\vert k\rangle\\
&&g^{(b,j,\eta,k)}(a,i,x^-,x)\ = \nonumber \\
&&\langle x|\theta(-x^-)e^{i{\partial^2\over 2k^-}x^-}(1-2l)_{ij}\delta^{ab}+\theta(x^-)[e^{i{D^2\over 2k^-}x^-}(1-2L)]^{ab}_{ij}+\epsilon(x^-)[\Delta (1-l-L)]^{ab}_{ij}|k\rangle\,.\nonumber
\end{eqnarray}
Thus
\begin{eqnarray}
\partial^+g^{*\,p}_{\ x}&=&\delta(x^-)\langle p|2l-2L+(1-l-L)\Delta^\dagger\vert x\rangle \\
&&-{i\over 2p^-}\langle p\vert \theta(-x^-)e^{-i{\partial^2\over 2p^-}x^-}\partial^2(1-2l)+\theta(x^-)[e^{-i{D^2\over 2p^-}x^-}D^2(1-2L)]|x\rangle\,.\nonumber
\end{eqnarray}
In the expression above we have not written all the indices explicitly. 
We first check the orthonormality of the basis functions $g$:
\begin{eqnarray}
\int_xg^*(p)\partial^+g(k)&=&{1\over 2}\langle p\vert (1-2l)(1-2L)-(1-2L)(1-2l)+{1\over 2}(2-2l-2L)\Delta(2-2l-2L)\vert k\rangle\nonumber\\
&+&\langle p\vert \int _{x^-<0}\Big[e^{-ix^-[{\partial^2\over 2p^-}-{\partial^2\over 2k^-}]}{i\partial^2\over 2k^-}-{i\over 8 k^-}(2-2l-2L)\Delta^\dagger (1-2l)\partial^2e^{ix^-{\partial^2\over 2k^-}}\Big]\vert k\rangle\nonumber\\
&+&\langle p\vert \int _{x^->0}\Big[e^{-ix^-[{D^2\over 2p^-}-{D^2\over 2k^-}]}{iD^2\over 2k^-}-{i\over 8 k^-}(2-2l-2L)\Delta^\dagger (1-2L)D^2e^{ix^-{D^2\over 2k^-}}\Big]\vert k\rangle\nonumber\\
&=&\langle p\vert \Big[{1\over 1-{k^-\over p^-}+i\epsilon}-{1\over 1-{k^-\over p^-}-i\epsilon}\Big]\vert k\rangle\nonumber\\
&+&{1\over 2}\langle p\vert (1-2l)(1-2L)-(1-2L)(1-2l)+{1\over 2}(2-2l-2L)(\Delta-\Delta^\dagger)(2-2l-2L)\vert k\rangle\nonumber\\
&=&-i2\pi\delta({k^-\over p^-}-1)=-i2\pi\delta(\eta-\xi)\,.\nonumber
\end{eqnarray}

Now for the calculation of the overlap matrices
\begin{eqnarray}
\Theta(p,p^-,k,k^-)&=&i\int_{x^-,x}\partial^+g^*(p)f(k)=i\langle p|2l-2L+(1-l-L)\Delta^\dagger\vert k\rangle\\
&+&\int_{x^-<0}{1\over 2\pi^-}\langle p\vert \partial^2e^{-ix^-[{\partial^2\over 2p^-}-{\partial^2\over 2k^-}]}(1-2l)\vert k\rangle\nonumber\\
&+&\int_{x^->0}{1\over 2\pi^-}\langle p\vert D^2(1-2L)e^{-ix^-[{D^2\over 2p^-}-{\partial^2\over 2k^-}]}\vert k\rangle\nonumber\\
&=&i\langle p|2l-2L+(1-l-L)\Delta^\dagger\vert k\rangle\nonumber\\
&+&{i\over 1-{p^-\over k^-}-i\epsilon}\langle p|(1-2l)\vert k\rangle-i\langle p\vert(1-2L)
{1\over 1-{\partial^2\over D^2}{p^-\over k^-}-i\epsilon}\vert k\rangle\,.\nonumber
\end{eqnarray}
By inspection we see that to get the matrix $\Phi$ we need to substitute $k^-\rightarrow -k^-$ and also $k\rightarrow -k$. Thus we have
\begin{eqnarray}
\Phi(p,p^-;-k,k^-)&=&i\langle p|2l-2L+(1-l-L)\Delta^\dagger\vert k\rangle\nonumber\\
&+&{i\over 1+{p^-\over k^-}-i\epsilon}\langle p|(1-2l)\vert k\rangle-i\langle p\vert(1-2L)
{1\over 1+{\partial^2\over D^2}{p^-\over k^-}-i\epsilon}\vert k\rangle\,.\nonumber
\end{eqnarray}
We have been a little cavalier about the non-commuting factors $D$ and $\partial$ in the above expressions. It is clear by construction that all factors of $D$ come from the function $g^*$. Thus all the  above expressions should be understood as ordered such that all factors $D$ are to the left of factors $\partial$.

\section{Appendix - Perturbative expansion.}
\label{sec:C}
Our aim in this appendix is to show explicitly that the $O(Y^2)$ terms discussed in Section 3 have  their exact counterpart in perturbative calculation. To do this we have to expand our expression for the matrix $K$ to first order in $j$. Our discussion will be for the wave function rather than for the $S$- matrix, but the extension is straightforward.

We thus consider the expansion of the operator $\Lambda$ to $O(j)$. This is straightforward and we give some details below. In all the expressions below the field $b$ has to be understood as $b^a_i={\partial_i\over \partial^2}j^a$. 
We start by expanding the matrix $\Phi$
\begin{equation}
\Phi(p,\eta;-k,\xi)=i\langle p\vert 2(l-L)+(1-l-L)\Delta^\dagger\vert k\rangle+{i\over 1+e^{\xi-\eta}}
\langle p\vert (1-2l)\vert k\rangle-i\langle p\vert (1-2L){1\over 1+{\partial^2\over D^2}e^{\xi-\eta}}\vert k\rangle\,.
\end{equation}
We have neglected the $i\epsilon$ terms in this expressions since the poles are never reached for physical values of rapidity.
\begin{equation}
2(l-L)^{ab}_{ij} = 2f^{cab}\left(\partial_i {1\over{\partial^2}}b^c_j + b^c_i{1\over{\partial^2}}\partial_j - 
\partial_i {1\over{\partial^2}}\left\{\partial_k, b^c_k\right\}{1\over{\partial^2}}\partial_j\right)\,.
\end{equation}
Further
\begin{equation}
\Delta^{\dagger ab}_{kj}=f^{cab}\partial_k \frac{1}{\partial^2} \left[\partial_l, b^c_l\right] \frac{1}{\partial^2} \partial_j-2f^{cab}
b^c_k \frac{1}{\partial^2}\partial_j\,.
\end{equation}
And
\begin{equation}
(1-l-L)^{ad}_{ij}\Delta^{\dagger db}_{kj}=f^{cab}\left\{-2b^c_i\frac{1}{\partial^2}\partial_j+\partial_i\frac{1}{\partial^2}b^c_k\partial_k\frac{1}{\partial^2}\partial_j+3\partial_i\frac{1}{\partial^2}\partial_kb^c_k\frac{1}{\partial}\partial_j\right\}\,.
\end{equation}
Thus the first term in $\Phi$ is
\begin{equation}
2(l-L)^{ab}_{ij}+(1-l-L)^{ad}_{ij}\Delta^{\dagger db}_{kj}=f^{cab}\left(2\partial_i\frac{1}{\partial^2}b^c_j + \partial_i\frac{1}{\partial^2}\left[\partial_k,b^c_k\right]\frac{1}{\partial^2}\partial_j\right)\,.
\end{equation}
The second term in $\Phi$ is of order one, while for the third term we need
\begin{eqnarray}
(1-2L)^{ad}_{ij}\left[\frac{1}{ 1 + \frac {\partial^2} {D^2} e^ {\xi-\eta} }\right]^{db} 
&=& \frac {1} { 1+ e^{\xi-\eta} } \left( \delta_{ij} -  2\partial_i \frac{1} {\partial^2} \partial_j \delta^{ab} \right)\nonumber\\
&&- \frac {f^{cab}} { 1+ e^{\xi-\eta} }  \left(  \delta_{ij} -  2\partial_i \frac{1} {\partial^2} \partial_j   \right)  \left\{\partial_k,b^c_k\right\}  \frac {1} {\partial^2}  \frac{e^{\xi-\eta}}{1+e^{\xi-\eta}}\nonumber\\
&&- f^{cab}\left( \partial_i\frac{1}{\partial^2}\left\{\partial_k,b^c_k\right\}\frac{1}{\partial^2}\partial_j -\partial_i\frac{1}{\partial^2}b^c_j -b^c_i\frac{1}{\partial^2}\partial_j\right)\,.
\end{eqnarray}
Thus finally
\begin{eqnarray}
\Phi^{ab}_{ij}&=&if^{cab}\left\{2\partial_i\frac{1}{\partial^2}b^c_j+\partial_i\frac{1}{\partial^2}\left[\partial_k,b^c_k\right]\frac{1}{\partial^2}\partial_j\right\}\nonumber\\
&&+if^{cab}\left(\partial_i\frac{1}{\partial^2}\left\{\partial_k,b^c_k\right\}\frac{1}{\partial^2}\partial_j - \partial_i\frac{1}{\partial^2}b^c_j-b^c_i\frac{1}{\partial^2}\partial_j\right)\nonumber\\
&&+if^{cab}\frac{1}{1+e^{\xi-\eta}}\left(\delta_{ij}-2\partial_i\frac{1}{\partial^2}\partial_j\right)\left\{\partial_k,b^c_k\right\}\frac{1}{\partial^2}\frac{e^{\xi-\eta}}{1+e^{\xi-\eta}}\,.\nonumber\\
\end{eqnarray}
The expansion of matrix $\Theta$ is trivial, since we only need its leading order term
\begin{equation}
\Theta^{ab}_{ij}=\left(\delta^{ab}_{ij}-2\partial_i\frac{1}{\partial^2}\partial_j\delta^{ab}\right)\left(\frac{i}{1-e^{\xi-\eta}-i\epsilon}-\frac{i}{1-e^{\xi-\eta}+i\epsilon}\right)=-\left(\delta^{ab}_{ij}-2\partial_i\frac{1}{\partial^2}\partial_j\delta^{ab}\right)\delta(\xi-\eta)\,.
\end{equation}
This yields
\begin{eqnarray}\label{la}
\Lambda^{ab}_{ij}(x,\eta;y,\xi)&=&\int_\zeta \Theta^{-1 ad}_{ik}(x,\eta,z,\zeta)\Phi^{db}_{kj}(z,\zeta,y,\xi)\nonumber\\
&=&if^{cab}\langle x\vert\left\{-\partial_i\frac{1}{\partial^2}b^c_j-b^c_i\frac{1}{\partial^2}\partial_j\right\}\vert y\rangle\nonumber\\
&+& if^{cab}\frac{e^{\xi-\eta}}{(1+e^{\xi-\eta})^2}\delta_{ij}\langle x\vert\left\{\partial_l,b^c_l\right\}\frac{1}{\partial^2}\vert y\rangle\,.\nonumber\\
\end{eqnarray}
In this calculation we are only interested in terms which contribute $O(Y^2)$ when $\Lambda$ is integrated over both rapidities. The second term in eq.(\ref{la}) exponentially decreases for $\xi-\eta\rightarrow\pm\infty$ and thus does not contribute. We can then limit ourselves to
\begin{equation}
\Lambda^{ab}_{ij}(x,\eta;y,\xi)\approx\lambda^{ab}_{ij}(x,y)\equiv -igf^{cab}\left\{\partial_i(x)\frac{1}{\partial^2}(x,y)b^c_j(y)+b^c_i(x)\frac{1}{\partial^2}(x,y)\partial_j(y)\right\}
\end{equation}
where we have also restored the coupling constant, omitted from the previous formulae for simplicity.
We now concentrate on the expansion of the evolved wave function
\begin{equation}
|\Psi\rangle=e^{i\int_{z,\epsilon}b^a_i(z)\left(a^a_i(z,\zeta)+a^{\dagger a}_i(z,\zeta)\right)}   
e^{-\frac{1}{2}\int_{x,y,\eta,\xi}a^{\dagger b}_k(x,\eta)\,\Lambda^{bc}_{kj}(x,y,\eta,\xi)\,a^{\dagger c}_j(y,\xi)}|0\rangle\,.
\end{equation}
Expanding in the Fock basis we find
\begin{eqnarray}
|\Psi\rangle&=&\left[1-Y\frac{1}{2}\int_{x,y}{b^a_i(x)b^a_i(y)}\right]|0\rangle\nonumber\\
&+&\left[i\int_{x,\eta}{b^a_i(x)a^{\dagger a}_i(x,\eta)}-ig{Y\over 2\pi}\int_{x,y,\eta}{b^a_i(x)\lambda^{ad}_{ij}(x,y)
a^{\dagger d}_j(y,\eta)}\right]|0\rangle
\nonumber\\
&+&\left[-\frac{g}{2}\int_{x,y,\eta,\xi}{\lambda^{uv}_{ij}(x,y)a^{\dagger u}_i(x,\eta)a^{\dagger v}_j(y,\xi)}-\frac{1}{2}\int_{x,y,\eta,\xi}{b^a_i(x)b^b_j(y)a^{\dagger a}_i(x,\eta)a^{\dagger b}_j(y,\xi)}\right]|0\rangle\nonumber
\end{eqnarray}
Hence the one gluon component is  can be written as
\begin{equation}\label{oneg}
\langle a,i, x,\eta|\Psi\rangle =ib^a_i(x) +g{Y\over 2\pi}f^{abc}b^b_i(x)\int_y\frac{\partial_i}{\partial^2}(x,y) b^c_j(y)=ig
{\partial_i\over\partial^2}j^a +g^3{Y\over 2\pi}f^{abc}{\partial_i\over\partial^2}(x,z)j^b(z)\frac{1}{\partial^2}(x,y) j^c(y)
\end{equation}
where now the charge $j$ is normalized so that the charge of a single gluon is of order unity.
The two gluon component of the wave function is
\begin{eqnarray}\label{twog}
&&\langle a,i,x,\eta;b,y,j,\xi\vert \Psi\rangle
=\,-g^2\int_u \partial_i\frac{1}{\partial^2}(x,u)j^a(u)\partial_j\frac{1}{\partial^2}(y,v)j^b(v)\nonumber\\
&&\ \ \ \ \ \ \ \ \ \ \ \ \ 
+\,ig^2 f^{abc}\int_u\left\{\partial_i\frac{1}{\partial^2}(x,y)\partial_j\frac{1}{\partial^2}(y,u)j^c(u)-\partial_j\frac{1}{\partial^2}(y,x)\partial_i\frac{1}{\partial^2}(x,u)j^c(u)\right\}\,.
\end{eqnarray}
These expressions have a very simple diagrammatic interpretation. 

The diagrams contributing to the one gluon state components are depicted in Fig.4 while to the two gluon components in Fig. 5. Clearly the first diagram on Fig.4 represents the first term in eq.(\ref{oneg}) and the second diagram - the second term in eq.(\ref{oneg}). The second term has an extra power of $Y$ due to the additional loop integral. Thus when calculating the $S$-matrix, the square of the first term is of order $O(Y)$, while the cross term is of order $O(Y^2)$. This is precisely the term we have encountered in our calculation in the body of the paper. 
\FIGURE{\centerline{\epsfig{file=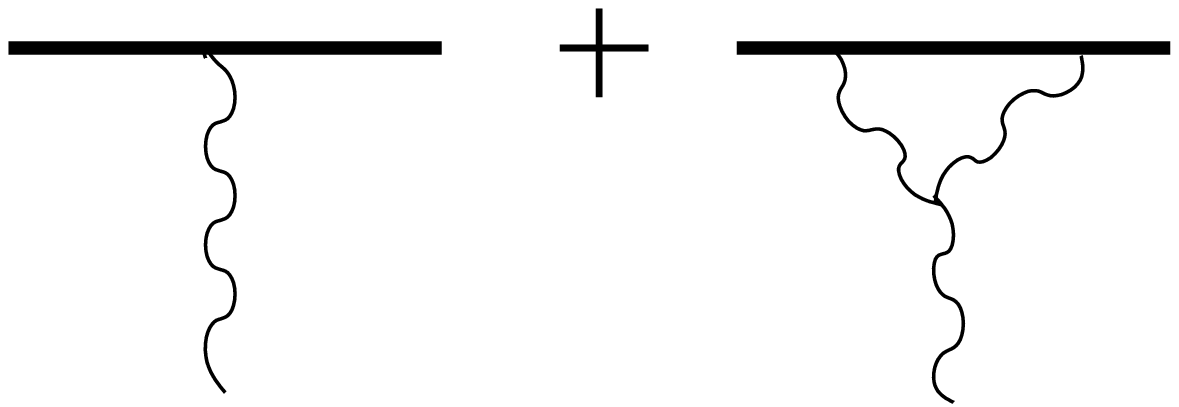,width=70mm}}
\caption{\it Diagrams contributing to the one gluon component of the wave function in perturbation theory.}
\label{fig4}
}

Similarly Fig.5 represents the two terms in eq.(\ref{twog}). Both terms here when squared, give contribution of $O(Y^2)$ to the $S$ matrix.

\FIGURE{\centerline{\epsfig{file=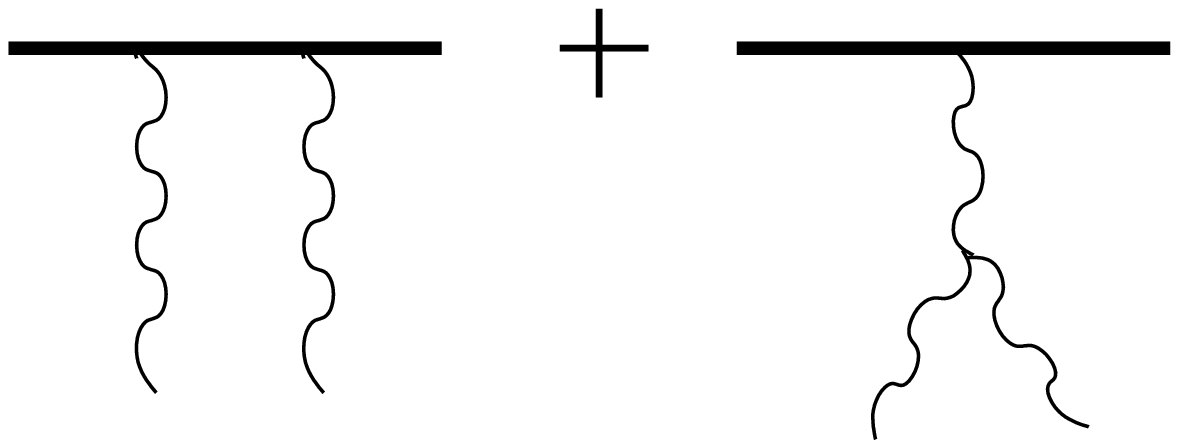,width=70mm}}
\caption{\it Diagrams contributing to the two gluon component of the wave function in perturbation theory.}
\label{fig5}
}


%
\end{document}